\documentclass[10pt, pre, amsmath, onecolumn, showpacs, superscriptaddress]{revtex4-1}
\usepackage{amsmath}
\usepackage{amssymb}
\usepackage[mathscr]{eucal}

\usepackage{graphicx}

\usepackage{color}
\definecolor{dgreen}{rgb}{0,0.7,0}

\newcommand{\dd}{\mathrm{d}}
\newcommand{\ee}{\mathrm{e}}
\newcommand{\ci}{\mathrm{j}}
\newcommand{\Erfc}{\mathrm{erfc}}
\newcommand{\Sign}{\mathrm{Sign}}
\newcommand{\Erf}{\mathrm{erf}}
\newcommand{\Erfi}{\mathrm{erfi}}
\newcommand{\Ei}{\mathrm{Ei}}
\newcommand{\gammae}{\gamma_\mathrm{E}}

\newcommand{\cA}{\mathcal{A}}
\newcommand{\cY}{\mathcal{Y}}
\newcommand{\cH}{\mathcal{H}}
\newcommand{\cC}{\mathcal{C}}
\newcommand{\cD}{\mathcal{D}}
\newcommand{\cV}{\mathcal{V}}

\newcommand{\zt}{\tilde{z}}
\newcommand{\zh}{\hat{z}}

\begin{document}
\title{Correlation and fluctuation in Random Average Process on an infinite line with a driven tracer 
}
\author{J. Cividini$^1$, A. Kundu$^{2}$, Satya N. Majumdar$^3$ and D. Mukamel}
% \author{J. Cividini, A. Kundu, S. N. Majumdar and D. Mukamel}
\affiliation{Department of Physics of Complex Systems, Weizmann Institute of Science, Rehovot 76100, Israel \\
$^2$International center for theoretical sciences, TIFR, Bangalore - 560012, India\\
$^3$LPTMS, CNRS, Univ. Paris-Sud, Universit\'e Paris-Saclay, 91405 Orsay, France.}
\date{\today}

\begin{abstract}
We study the effect of single biased tracer particle in a bath of other particles performing the random average process (RAP) on an infinite line. 
We focus on the large time behavior of the mean and the fluctuations of the positions of the particles and also the correlations among them. 
In the large time $t$ limit these quantities have well-defined scaling forms and grow with time as $\sqrt{t}$.
% Because of the biased motion of the tracer, the average particle density profile gets modulated. 
% We also compute the modulation in the particle density created by the biased tracer.
% The time dependence of the mean particle density profile has also been studied. 
% Since the particles can not overtake each other, they are subjected to same caging effect as in other non-overtaking particle systems in one dimension. 
% As a result these particles become strongly correlated. 
% In this paper we study the two-point correlations of the positions and find that they posses scaling forms in large $t$ limit. 
A differential equation for the scaling function associated with the correlation function is obtained and solved 
perturbatively around the solution for a symmetric tracer. Interestingly, when the tracer is 
totally asymmetric, further progress is enabled by the fact that the particles behind of the tracer do not affect the motion of the particles in front of it,
which leads in particular to an exact expression for the variance of the position of the tracer. Finally, the variance and correlations 
of the gaps between successive particles are also studied. Numerical simulations support our analytical results.
\end{abstract}

\maketitle

\section{Introduction}
\noindent
The motion of non-overtaking particles in narrow channels are known as single-file diffusion.
In such one dimensional geometry the motion of any particle is hemmed by its neighbors. As a 
result the particles cannot bypass each other and keep their initial order the same over time. 
Study of such restricted motion of particles have been started by 
Harris \cite{Harris-65} and Jepsen \cite{Jepsen-65}. They showed that when the particles evolve according 
to Hamiltonian dynamics the mean squared displacement (MSD) of a tagged particle [also called tracer particle (TP)] grows diffusively, 
whereas for Brownian particles the MSD of a TP grows subdiffusively. Recently, several experiments have been able to observe TP diffusion by passive 
microrehology in zeolites, transport of colloidal particles or charged spheres in narrow circular channels \cite{Gupta-95, Kulka-96, 
Hahn-96, Wei-00, Meersmann-00, Lin-05, Lutz-04}.  
Such experimental evidences have generated 
a great revival of interest in tagged particle diffusion.
Many different results regarding the tagged particle diffusion have been reported for various systems with differently organized dynamics 
\cite{Percus-74, Beijeren-83, Arratia-83, Alexander-78, Majumdar-91, Lizana-09, Barkai-09, Kollmann-03, Gupta-07, Rodenbeck-98,
 Barkai-10, Roy-13, Sabhapandit-07, Illien-13, Benichou-13, Krapivsky-14}. 
For example, in addition with the mean position and the MSD of the TP, 
probability distribution functions (PDF) associated with particle displacements have been studied as well \cite{Rodenbeck-98, Barkai-09, Barkai-10, Lizana-09, 
Krapivsky-14, Hegde-14, Illien-13, Sabhapandit-15}. 

One of the simplest system where tagged particle diffusion has been studied in detail is the simple exclusion process (SEP).
This process is usually defined on a one dimensional lattice, where each lattice site is occupied 
by one hardcore particle or it is empty. In every small time interval $dt$, each particle moves to the neighboring site 
on the right with probability $\alpha dt$ and on the left with probability $\beta dt$ iff the target site is empty. 
The hardcore interaction among the particles plays dramatic role 
in the long time asymptotic growth of MSD of a tagged particle. In the absence of bias ($\alpha = \beta = 1/2$), 
the mean squared fluctuation of the displacement of a TP 
grows subdiffusively as $\sim A_0 \sqrt{t}$ for large $t$ where the prefactor $A_0$ is given explicitly in terms of particle density $\rho_0$ as 
$A_0=\frac{1-\rho_0}{\rho_0}\sqrt{\frac{2}{\pi}}$ \cite{Harris-65, Arratia-83, Alexander-78}. On the other hand when there is non-zero bias $\alpha \neq \beta$, 
the MSD grows diffusively as $\sim (\alpha-\beta)(1-\rho_0) t$ for large $t$ \cite{Demasi-85, Kutner-85}. 

Similar results have also been proved for another interesting interacting and widely studied 
many particle system called the \textit{Random Average Process} (RAP) 
 first introduced by Ferrari and Fontes \cite{Ferrari-98}.
In RAP particles move on a one dimensional continuous line in contrast to SEP where hardcore particles move on a lattice. Each particle moves to 
the right (left) by a \textit{random fraction} of the space available until the next nearest particle on the right (left) with some rate $\alpha$ ($\beta$).
 %probability $pdt$ ($qdt$).
Thus the jumps in either direction is a random fraction $\eta$ of the \textit{gap} to the nearest particle in that direction where the random number 
$\eta \in [0,1)$ is chosen from some distribution $R(\eta)$. As a result the particles in RAP also 
never overtake each other keeping their initial order unchanged over time as in other 
single-file motion. The RAP appears in a variety of problems like the force propagation in granular media \cite{Coppersmith-96, Rajesh-00}, 
in porous medium equation \cite{Feng-96}, 
in models of mass transport \cite{Krug-00, Rajesh-00}, models of
voting systems \cite{Melzak-76}, models of wealth distribution \cite{Ispolatov-98} and in the generalized Hammersley process \cite{Aldous-95}. 
In the unbiased ($\alpha=\beta$) case the MSD of a TP in RAP also grows subdiffusively 
as $\sim A \sqrt{t}$ whereas it diffuses as $\sim D~t$ in the globally uniform bias ($\alpha \neq \beta$) case for large time \cite{Rajesh-01}. 
The constants $A$ and $D$ in the prefactors are computed exactly in terms of particle density $\rho_0$ and the moments of the jump distribution $R(\eta)$ \cite{Rajesh-01}.
% by $A_{SRAP}=2\rho_0^{-2}\sqrt{p\mu_1/\pi}~\mu_2/(\mu_1-\mu_2)$ and $D_{ARAP}=\rho_0^{-2}(p-q)\mu_1\mu_2/(\mu_1-\mu_2)$ where $\mu_k$ is the $k$-th moment 

In single-file motion, the movements of individual particles become in general strongly correlated because any large progressive displacement 
of a given particle in one direction also necessarily requires large displacements of more and more other particles in the same direction. 
In the context of RAP, such correlation between positions of two tagged particles have been computed explicitly in terms of 
their label separation $r$ and time $t$. If $x_i(t)$ represents the position of the $i$-th particle, then for large $t$, the correlation function
 $c_{i,j}(t)= \langle [x_i(t) -\langle x_i\rangle(t)][x_j(t) -\langle x_i\rangle(t)] \rangle$ is given by the following scaling form \cite{Rajesh-01}, 
\begin{eqnarray}
c_{i,j}^\mathrm{gb}(t)&=& \rho_0^{-2}\frac{\mu_2}{\sqrt{2\pi}(\mu_1-\mu_2)}~\sqrt{2\mu_1(\alpha+\beta)t}
~g\left( \frac{i-j}{\sqrt{2\mu_1(\alpha+\beta)t}}\right),~~\text{where,}~~\label{C_r-RM} \\ 
&&~~~~~~~g(u)=e^{-\frac{u^2}{2}} - \sqrt{\frac{\pi}{2}} ~|u|~\Erfc\left( \frac{|u|}{\sqrt{2}}\right), \label{g_u_0}
\end{eqnarray}
where $\mu_1$ and $\mu_2$ are the first and second moments, respectively, of the jump distribution $R(\eta)$ and the superscript `gb' indicates `global bias'.
Note that the scaling 
function $g(u)$ is independent of the parameters $\alpha$ and $\beta$, \textit{i.e.}
independent of the global bias.
%Note that the correlations between positions in the large time limit are described by the same scaling function $g(u)$ no matter whether the system is unbiased or 
%globally and uniformly biased. 
Naturally a question arises : what happens if the system is locally biased instead of globally biased ? More precisely, 
if a single particle in RAP 
moves asymmetrically while all others are moving symmetrically, how does the MSD grow with time ?
How correlated are the positions of two particles ? 
% Does the correlation function also have a scaling form in large time limit ? If so then what is it explicitly ?
In this paper we address these questions.

Motion of single driven tracer particle (DTP) in the pool of other non-driven interacting ( hardcore interaction with the tracer particle and among others ) 
particles have been studied in various contexts. In experimental studies, single driven tracer in quiescent media have been 
used to probe rheological properties of complex media such as DNA \cite{Gusche-08},
polymers \cite{Krager-09}, granular media \cite{Candelier-10, Pesic-12} or colloidal crystals \cite{Dullens-11}. 
Some practical examples of biased tracer are a charged
impurity being driven by applied electric field or a colloidal particle being pulled by optical tweezer in presence of other
colloid particles performing random motion. On the theoretical side, situations have been considered where 
the surrounding medium is a Symmetric Simple Exclusion 
Process (SSEP) and the tagged particle is a hard-core tracer driven towards a preferred direction. 
%In dimensions larger than one, in the stationary state the tracer acquires a finite velocity. In the tracer frame 
%the perturbation of the density profile becomes stationary\greenw{RTEF} and consists in a depleted 'cone' at the 
%back of the tracer and denser zones at the front and on the sides. In this system interesting effects have been 
%evidenced, such as the fact that the perturbation of the density profile decays algebraically at the back of the 
%tracer\greenw{REF}. It has also been shown that the displacement distribution of the tracer converges to a Gaussian 
%at large times, but the variance may grow anomalously depending on the geometry\greenw{REF}.
In this context the effect of the biased tracer has been quantified in terms of both the tracer motion and 
the perturbation of the density profile~\cite{Burlatsky-92,Burlatsky-96,Landim-98,Benichou-99,Coninck-97,Benichou-01,Benichou-00}. 
Contrary to what happens in higher dimensions~\cite{Coninck-97,Benichou-01,Benichou-00}, 
the velocity of the tracer moving in a 1D SSEP vanishes~\cite{Burlatsky-92,Burlatsky-96,Landim-98,Benichou-99}. 
It has been shown theoretically that the perturbation of the density field of the bath particles generically consists in a denser region in the front  
and a depleted region at the back of it as expected intuitively. The amplitude of the difference between the density profiles in the biased and the unbiased case, 
decays to zero exponentially as one goes far from the driven tracer on both sides in 1D. However in higher dimensions such decay is dependent on the direction along which
one moves away from the driven tracer \cite{Burlatsky-92,Burlatsky-96,Landim-98,Benichou-99,Coninck-97,Benichou-01,Benichou-00}.
% The perturbation of the density field of the bath particles generically consists in a denser region in front of 
% the tracer whose amplitude decays exponentially and of a depleted region at the back, that has been shown to exhibit 
% algebraic decay in dimensions larger than one~\cite{Coninck-97,Benichou-01,Benichou-00}.

In the next section we define the model and present the summary of our results.
% There exists almost no such results in the context of RAP with single driven tracer particle. In this paper we derive new 
% results concerning the fluctuation and correlations of the positions of the particles performing RAP.
% The detailed study of the RAP is important because 

% {\it The detailed study of the RAP is important
% since it has shown up either directly or in disguise in a
% variety of problems including traffic models [4], models
% of mass transport [5], models of force fluctuation in bead
% packs [6], models of voting systems [7,8], models of wealth
% distribution [9] and the generalized Hammersley process
% [10]. Like the simple exclusion process, some aspects of
% the RAP are analytically tractable [4,5,11]. }

\section{Model definition and summary of the results}
\label{Model}
\begin{figure}[!ht]
	\begin{center}
		\includegraphics[width=0.6\textwidth]{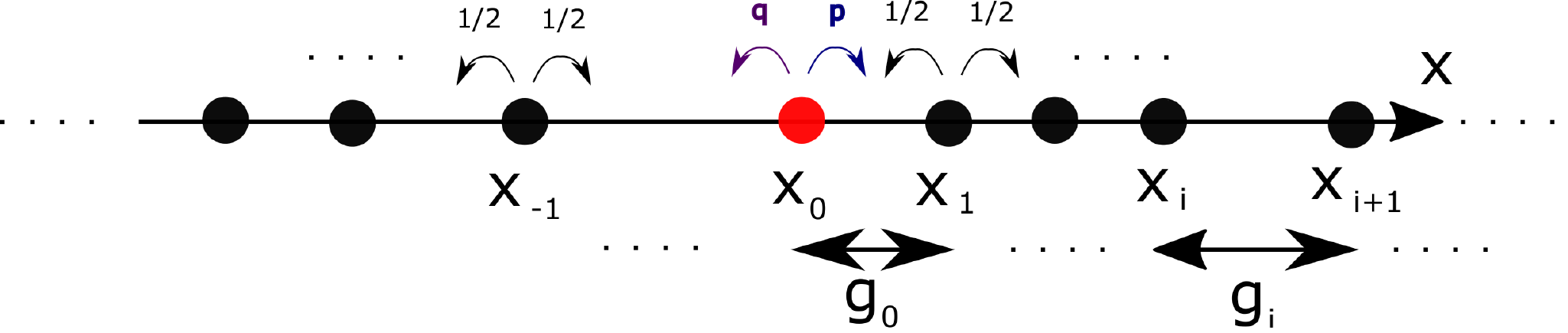} 
	\end{center}
	\caption{\small Schematic diagram of the RAP with a driven tracer particle 
                   ( $0$-th particle colored in red ) on an infinite line. The variable $x_i$ represents the position of the 
                 $i$-th particle and $g_i=x_{i+1}-x_{i}$ represents the gap between the $(i+1)$-th and $i$-th particles. The driven tracer particle hops to the left with 
                 rate $p$ and to the right with rate $q$, whereas all other particles ($i\neq 0$) hop to the left or to the right with the same rate $1/2$. }  
	\label{fig:scheme}
\end{figure}
\noindent
We consider an infinite number of particles occupying an infinite line with density $\rho_0$. 
Without any loss of generality, we label the driven tracer particle as the $0$th particle and then 
label other particles according to their positional order with respect to the tracer particle from $-\infty$ to $\infty$.
% Each particle is labeled with an integer $i=-\infty,\ldots,\infty$ ,
% say from left to right (see fig. \ref{fig:scheme}). We consider the particle labeled $0$ as the driven tracer particle (DTP). 
Let us denote the positions of the particles at time $t$ by $x_i(t) \in \mathbb{R}$ for $i \in \mathbb{Z}$.
Initially \emph{i.e.} at $t=0$ the particles are arranged according to the following fixed configuration
\begin{equation}
\label{eq:xini}
x_i(0)=\rho_0^{-1} i,~~~i =-\infty,...,-1,0,1,...,\infty.
\end{equation}
Hence all the averages $\langle .... \rangle$ in this paper are taken over stochastic evolution. 
The dynamics of the particles are given as follows. 
In an infinitesimal time interval $t$ to $t+dt$, any particle (say $i^{\text{th}}$) other than the DTP, jumps from $x_i(t)$, either to the right 
or to the left with probability $dt/2$  
and with probability $(1-dt)$ it stays at $x_i(t)$. The DTP jumps from $x_{0}(t)$ to the right with probability $pdt$, to the left with probability $qdt$ 
and does not jump with probability $(1-(p+q)dt)$.
The amount of jump, either to the right or to the left, made by any particle is a random fraction 
of the space available between the particle and its neighboring particle to the right or to the left.
For example, the $i^{\text{th}}$ particle jumps by $\eta_i^r [x_{i+1}(t) - x_i(t)]$ to the right and 
by $\eta_i^l [x_{i-1}(t) - x_i(t)]$ to the left. The random variables $\eta_i^{r,l}$ are independently chosen from the interval $[0,~1)$ and each is distributed 
according to the same distribution $R(\eta)$, with moments 
\begin{equation}
\mu_k = \int_{\eta=0}^{1} \eta^k R(\eta) d\eta.
\end{equation} 
The time evolution of the positions $x_i(t)$’s can be written as,
\begin{eqnarray}
x_i(t+dt)= x_i(t) &+& \sigma_r^i \eta_i ~[x_{i+1}(t)-x_i(t)] + \sigma_l^i \eta_i~[x_{i-1}(t)-x_i(t)]
\label{dyna-x}
\end{eqnarray}
where the $\eta$ variables are independent and identically distributed according to $R(\eta)$. For $i \neq 0$, $\sigma^i_{r}$ and $\sigma^i_{l}$ are $1$ with probability $\frac{dt}{2}$
and $0$ otherwise. The random variable $\sigma^0_r$ is $1$ with probability $p dt$ and $0$ with probability $1- p dt$. Similarly, $\sigma^0_l$ 
is $1$ with probability $q dt$ and $0$ with probability $1- q dt$. Clearly, we see that all the particles are symmetrically moving except the $0$-th particle, which moves 
asymmetrically. In this paper we are mainly interested in the effect of this asymmetric motion of TP on the fluctuations and the correlations among 
the positions of other particles.

% \begin{itemize}
%  \item 
We first look at the effect of the biased tracer on the time dependence of the average position $y_i(t)=\langle x_i(t) \rangle$, 
average gap $h_i(t)=\langle g_i(t) \rangle = \langle x_{i+1}(t)-x_i(t) \rangle$ and 
the average particle density $\rho(w,t)=\langle \sum_{i=-\infty}^\infty \delta[w-x_i(t)]~\rangle$ profile in section \ref{section:1pt}.
In the large time limit we find that these three quantities have the following scaling forms 
\begin{eqnarray}
 y_i(t) &=& \rho_0^{-1}~\sqrt{2\mu_1 t}~\cY \left(\frac{i}{\sqrt{2\mu_1 t}} \right) + o(\sqrt{t}), \nonumber \allowdisplaybreaks[4]\\
 h_i(t) &=& \rho_0^{-1}~\cH \left(\frac{i}{\sqrt{2\mu_1 t}} \right) + o(\sqrt{t}),  \allowdisplaybreaks[4] \\
\rho(w,t)&=& \Omega\left(\frac{w}{\sqrt{2 \mu_1 t}} \right)+ o(\sqrt{t}),\nonumber
\end{eqnarray}
where the index variables $i$ and the space variables $w$ are rescaled appropriately by time $t$. Explicit forms of these 
scaling functions are given in \eqref{eq:Ysol}, \eqref{eq:scalingH} and (\ref{pos-indx-rela}-\ref{den-scling}) respectively.
Here $o(\ell)$ represents contribution at order smaller than $\ell$.
Note that in the above three equations the density $\rho_0$ appears as an overall factor. This is because the dynamics is invariant under 
a rescaling of the position variables $x_i \rightarrow a x_i$. 
Consequently, we expect that $\rho_0$ will appear only as an overall factor in different average quantities \textit{e.g.} in mean positions, 
in correlation functions etc.
Since the gap variables $g_i(t)$s are equal to $x_{i+1}(t)-x_i(t)$, the scaling functions $\cY(x)$ and $\cH(x)$ associated with
$\langle x_i(t)\rangle$ and $\langle g_i(t) \rangle $ respectively, 
are related by $\cH(x)=\partial_x\cY(x)$.
We compute these scaling functions $\mathcal{Y}(x)$, $\cH(x)$ and also $\Omega(\xi)$ exactly in section \ref{section:1pt}
%  \ref{section:av-pos-sc}, \ref{subsection:av-gap-sc} and in \ref{subsection:av-den-sc}, respectively, 
and compare them with numerical measurements. For obvious reasons, we perform our numerical simulations on a 
ring of size $L$ with large $N$ number of particles. In all our simulations we consider $L=1$, $N=200$ (unless otherwise specified) and uniform jump distribution $R(\eta)=1$, whose moments are $\mu_k = \frac{1}{k+1}$. 
In the simulation, we observe that the late time growth of the average position $y_0(t)$ of the DTP changes from $\sim B_{line}~\sqrt{t}$ to linear growth $\sim B_{ring}~t$ 
as $t$ is increased. We compute the constants $B_{line}$ and $B_{ring}$ theoretically and compare them with numerical measurements. In particular, we find that the crossover between the line and the ring geometries 
can be captured through a nice crossover function which is given explicitly in \eqref{eq:phi} and plotted in fig. \ref{fig:crossover}.

% However, when time $t$ is not too large [$\ll O( \rho_0^2 L^2)$] 
% such that the particles moving on the ring have not yet realized the finiteness of the ring, it effectively acts as an infinite line. 
% As a result at earlier times, $y_0(t)$ grows as $\sim B_{line}~\sqrt{t}$, but as time 
% increases, the growth rate changes to $\sim B_{ring}~t$. We compute the constants $B_{line}$ and $B_{ring}$ exactly and study this crossover. 

% \item 
Next in section \ref{section:2pt} we study the pair position correlation function $c_{i,j}(t)=\langle x_i(t) x_j(t) \rangle -y_i(t) y_j(t)$ 
and the pair gap correlation function $d_{i,j}(t)=\langle g_i(t) g_j(t) \rangle -h_i(t) h_j(t)$.
%In case where all the particles are the same \emph{i.e.} hopping rate to the right and to the left are the same, say $1/2$, 
%many results concerning the two-point correlations of the positions have been derived by Rajesh and Majumdar \cite{Rajesh-01}. 
In the case where all particles hop symmetrically to the right and to the left with a rate, say, $1/2$, many results concerning the two-point correlations of the positions have been derived by Rajesh and Majumdar \cite{Rajesh-01}. 
In fact, in \cite{Rajesh-01} a more general situation have been considered where all the particles are identical in the sense that all of them have the same hopping rate $\alpha$ to the right and the same hopping rate $\beta$ to the left.
% ( different from $\alpha$ in general ) 
For this case, the two-point correlation function $c_{i,j} (t)$ has been computed.
In this translationally invariant case the correlation 
function $c_{i,j}(t)$ depends only on the label separation (or the initial separation) $r=|i-j|$ between the two particles and on time $t$. 
Moreover,in the large $t$ limit it was found that the correlation 
function has a scaling form in terms of the rescaled variable $\frac{r}{\sqrt{t}}$, see~(\ref{C_r-RM}).
On the other hand, in the model considered in the present paper the system is not translationally invariant as one particle ($0$-th particle) 
is driven and others are symmetrically moving. As a result, in our case the correlation 
function $c_{i,j}(t)$ depends on the indices $i$ and $j$ individually. Although, as we will later see, $c_{i,j}(t)$ in our case also has a scaling form :
\begin{equation}
 c_{i,j}(t)=\rho_0^{-2} \sqrt{2 \mu_1 t} ~\cC \left( \frac{i}{\sqrt{2 \mu_1 t}}, \frac{j}{\sqrt{2 \mu_1 t}}\right) + o(\sqrt{t}), \label{c-ij-sc}
\end{equation}
in terms of the rescaled variables $x=\frac{i}{\sqrt{2 \mu_1t}}$ and $y=\frac{j}{\sqrt{2\mu_1t}}$ where $\rho_0$ is the particle density and 
$\mu_1$ is the first moment of the jump distribution $R(\eta)$. 
In the beginning of sec. \ref{section:2pt} we numerically
verify that in large $t$ limit, $c_{i,j}(t)$ indeed has the scaling form \eqref{c-ij-sc}.
%  and $d_{i,j}(t)$ has scaling form \eqref{d-ij-sc} and \eqref{d-ii-sc}. 
Next inserting the form \eqref{c-ij-sc} in discrete evolution equation for $c_{i,j}(t)$ and taking large time limit, 
we obtain a differential equation for $\cC(x,y)$ in sec. \ref{C-diff-eq}. 
In section \ref{pertur-sol-C} we present a perturbative solution for $\cC(x,y)$ where we start with the following expansion 
\begin{equation}
 \cC(x,y) = \cC_0(x,y) + \frac{\epsilon}{2} \cC_1(x,y) + \frac{\epsilon^2}{4} \cC_2(x,y) + ...
\end{equation}
in powers of the drive strength $\epsilon =p-q$. As a result we get individual equations for each $\cC_i(x,y)$ with sources depending on lower order functions. 
One can in principle solve for each $\cC_i(x,y)$ separately. In this paper we compute $\cC_0(x,y)$ and $\cC_1(x,y)$ explicitly and compare them with numerical measurements. 
Some details of the computation of $\cC_1(x,y)$ have been left in Appendix \ref{section:appC1}. 

% \item 
The $q=0$ case is a special case as for this case the boundary conditions associated with the differential equation for $\cC(x,y)$ becomes simpler. This allows us to 
use the image method to solve $\cC(x,y)$ exactly in the first quadrant  $x \geq 0$ and $y \geq 0$. In particular, we compute the variance $\sigma_0^2(t)$ of the position of the driven tracer particle exactly 
as a function of time $t$. In sec. \ref{q_0-case} we prove, 
\begin{equation}
 \sigma_0^2(t)=\langle x_0^2(t) \rangle -\langle x_0(t) \rangle^2 
= \rho_0^{-2} \frac{\mu_2}{\mu_1-\mu_2}~\sqrt{\frac{2}{\pi}}~\frac{\sqrt{2}-1}{\sqrt{2}+1}~\sqrt{2\mu_1 t} + o(\sqrt{t}).
\end{equation}

% \item 
The two-point gap correlation function $d_{i,j}(t)$ is studied in sec. \ref{gap-corr}.  
Similar to position correlation function $c_{i,j}(t)$, the gap correlation $d_{i,j}(t)$ also supports scaling form under the same rescaling of indices 
: $x=\frac{i}{\sqrt{2 \mu_1t}}$ and $y=\frac{j}{\sqrt{2\mu_1t}}$. In particular we find that the diagonal $d_{i,i}(t)$
and the non-diagonal $d_{i,j}(t)$ gap correlations have different scaling forms in the large $t$ limit :
\begin{eqnarray}
\label{eq:scalingd0}
d_{i,j} &=& \frac{\rho_0^{-2}}{\sqrt{2 \mu_1 t}} \cD \left( \frac{i}{\sqrt{2 \mu_1 t}}, \frac{j}{\sqrt{2 \mu_1 t}}\right) + O(t^{-1}), \qquad i \neq j, \label{d-ij-sc} \\
d_{i,i} &=& \rho_0^{-2} \cV \left( \frac{i}{\sqrt{2 \mu_1 t}}\right) + \frac{\rho_0^{-2}}{\sqrt{2 \mu_1 t}} \left[ \cV_1 \left( \frac{i}{\sqrt{2 \mu_1 t}}\right)  
+ \cD \left( \frac{i}{\sqrt{2 \mu_1 t}}, \frac{i}{\sqrt{2 \mu_1 t}}\right) \right] + O(t^{-1}).\label{d-ii-sc}
\end{eqnarray}
In sec. \ref{gap-corr} we compute the scaling functions $\cV(x)$ and $\cV_1(x)$ exactly. 
      The scaling function $\cD(x,y)$ associated with the off-diagonal correlation function $d_{i,j}(t)$ can be obtained from $\cC(x,y)$ as they are related via  
     $\cD(x,y)=\partial_x\partial_y\cC(x,y)$. 
% \item 
Finally in sec. \ref{conclusion}, we conclude the paper.

% \end{itemize}
% In this paper we compute these scaling functions $\cC(x,y),~\cD(x,y),~\cV(x) $ and $\cV_1(x)$ analytically. 
% \begin{itemize}
%  \item 

% \item 

% \item 

% \end{itemize}

\section{Average position and particle density profile}
\label{section:1pt}
\noindent
When there is no biased tracer particle, the average positions of the particles remain the same as their initial positions \emph{i.e.} $y_i(t)=x_i(0)$. But in presence of 
biased tracer particle this will naturally not hold.
%  because the motion of the biased tracer affects the motion of other particles. 
In this section, we compute its effect on the average position $y_i(t)=\langle x_i(t) \rangle$, mean gap $h_i(t)=\langle g_i(t)\rangle = \langle x_{i+1}(t)-x_i(t) \rangle$ 
and the mean particle density $\rho(w,t)=\langle \sum_{i=-\infty}^\infty \delta[w-x_i(t)]~\rangle$ profile. 
\begin{figure}[!ht]
	\begin{center}
		\includegraphics[width=0.5\textwidth]{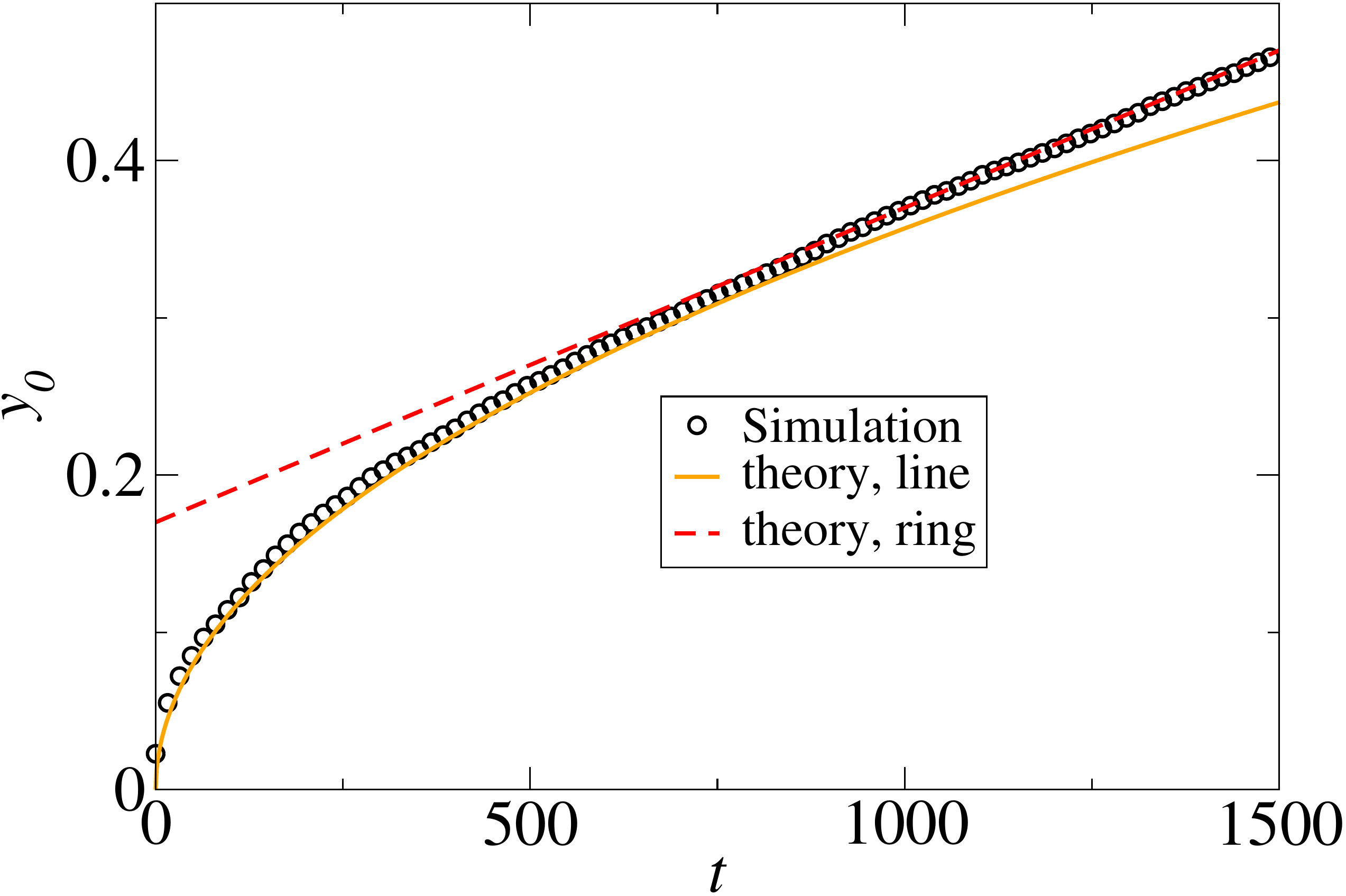} 
	\end{center}
	\caption{\small Average position $y_0(t)$ of the DTP as a function of time. Simulations are done on a ring of size $L=1$ with 
                 $N=50$ particles. The hopping rates of the DTP for this plot are $p=0.5$ and $q=0$. The red dashed line corresponds to \eqref{eq:z0ring} 
                 (shifted along y-axis) and the orange solid line corresponds to \eqref{eq:z0line}. Jump distribution is uniform \emph{i.e.} $R(\eta)=1$. }  
	\label{fig:yt}
\end{figure}

\subsection{Average positions : $y_i(t)=\langle x_i(t) \rangle$}
\label{subsection:Av-pos}
\noindent
Since the tracer particle ($0$-th particle) is driven in our model, it will induce an average motion of other particles 
in the direction of the drive. As a result their average 
positions $y_i(t)$ at time $t$ will grow from their initial positions $x_i(0)=\rho_0^{-1}~i$. When $p \neq q$, we would intuitively expect that the particles will 
acquire a velocity and hence their average positions will grow linearly with time $\sim t$. 
However, as we will shortly see, their positions grow as $\sim \sqrt{t}$ in the large time limit \emph{i.e} the velocity vanishes. 
On the other hand, if one looks at the motion of the particles on a finite ring then in the large $t$ limit the particles' velocity does not go to zero and their  
average positions grow linearly with time. To observe this crossover in the asymptotic growth of mean positions $y_i(t)$, 
we start with the motion of $N$ particles on a ring of size $L$ although originally our model is defined on an infinite line [see (\ref{dyna-x})]. In the end we take 
the following two limits $t\to \infty$ and $N =\rho_0 L \to \infty$ keeping the density $\rho_0=N/L$ fixed. We find that $t \to \infty$ and $N \to \infty$ do not 
commute. When $t\to \infty$ before $N\to \infty$, we find $y_i(t) \sim t$ whereas the opposite sequence of limits \emph{i.e.} first $N\to \infty$ then $t\to \infty$, 
yields $y_i(t) \sim \sqrt{t}$. We here emphasize that, the model on the ring is considered only in this section \ref{subsection:Av-pos} just to observe this crossover. 
In all other sections, we work with the original model (\ref{dyna-x}) defined on an infinite line.

Similar to \eqref{dyna-x}, one can write the dynamics of the particles on a ring \cite{Julien-15}, from which the evolution equation 
for the average positions $y_i(t)= \langle x_i(t) \rangle$ can be easily computed. It is however, convenient to work with the displacement variables 
$z_i(t) = y_i(t) - y_i(0)$. One writes the evolution equations for $z_i(t)$s as
\begin{equation}
\label{eq:evoziring}
\begin{cases}
& \dot{z}_0 = \mu_1 p (z_1 - z_0) + \mu_1 q (z_{N-1}-z_0) +\mu_1 (p-q) \rho_0^{-1}, \\
& \dot{z}_i = \frac{\mu_1}{2}(z_{i+1} - 2 z_i + z_{i-1}), \qquad i=1,\ldots, N-2\\
& \dot{z}_{N-1} = \frac{\mu_1}{2}(z_{0} - 2 z_{N-1} + z_{N-2}),
\end{cases}
~~~\text{where}~~~z_i(0)=0.
\end{equation}
and $\dot{z}=dz/dt$.
We solve equations~\eqref{eq:evoziring} by taking joint Fourier-Laplace transforms. Rescaling time by $\tau= \mu_1t/2$, we define the Laplace transform 
\begin{equation}
\label{eq:defltz}
\zt_i(s) = \int_{0}^{\infty} \ee^{-\tau s} z_i(\tau) \dd \tau
\end{equation}
and the joint Fourier-Laplace transform 
\begin{equation}
\label{eq:deffltz}
\zh_k(s) = \sum_{i=0}^{N-1} \ee^{-\frac{2 \pi \ci k i}{N}} \zt_i(s),
\end{equation}
where $\ci^2 = -1$. 
The inverse Fourier transform is given by
\begin{equation}
\label{eq:invfltz}
\zt_i(s) = \frac{1}{N} \sum_{k=0}^{N-1} \ee^{\frac{2 \pi \ci k i}{N}} \zh_k(s).
\end{equation}
After performing joint Fourier-Laplace transformation on both sides of \,\eqref{eq:evoziring} we get
\begin{eqnarray}
\label{eq:zksU}
\zh_k(s) &=& \frac{2 (p-q) \rho_0^{-1}}{s \lambda_k(s)} + \frac{U(s)}{\lambda_k(s)},~~~\text{where}, \\
\lambda_k(s) &=& s + 4 \sin^2 \left( \frac{\pi k}{N} \right), 
\end{eqnarray}
and $U(s) = (2p-1) (\zt_1(s)-\zt_0(s)) + (2q-1) (\zt_{N-1}(s)-\zt_0(s))$. Determining $U(s)$ self consistently and performing inverse Fourier transform we get 
\begin{equation}
\label{eq:zls}
\zt_i(s) =  \frac{\frac{1}{N} \sum_{k=0}^{N-1} \frac{\ee^{\frac{2 \pi \ci k i}{N}}}{\lambda_k(s)}}{1-\frac{1}{N} \sum_{k=0}^{N-1} \frac{(2p-1) 
(\ee^{\frac{2 \pi \ci k}{N}}-1) + (2q-1) (\ee^{\frac{-2 \pi \ci k}{N}}-1)}{\lambda_k(s)}} \frac{2 (p-q) \rho_0^{-1}}{s}.
\end{equation}
Now taking inverse Laplace transform of $\zt_i(s)$ one can in principle find $z_i(t)$ for any $t$. However 
we are interested in the long time limit, which is equivalent to studying the $s \rightarrow 0$ limit of~\eqref{eq:zls}. 
Here two cases arise depending on whether we take the thermodynamic limit ( $N\to \infty$ keeping $\rho_0=N/L$ fixed ) before $s \to 0$ or after.
Let us focus on the average displacement of the DTP ($i=0$), separately for these two cases :

\begin{itemize}
 \item[] (a) If $N$ is kept finite and $s \rightarrow 0$, the sum at the numerator is expected to be dominated by the $k=0$ term, as $\lambda_0(s) = s$. 
The sum at the denominator converges to a finite value in $s \to 0$ limit : 
\begin{equation}
\label{eq:sumdenom}
\frac{1}{N} \sum_{k=0}^{N-1} \frac{(2p-1) (\ee^{\frac{2 \pi \ci k}{N}}-1) + (2q-1) (\ee^{\frac{-2 \pi \ci k}{N}}-1)}{\lambda_k(0)} = - (p + q - 1) \frac{N-1}{N}.
\end{equation}
Hence in $s \to 0$ limit, 
\begin{equation}
\label{eq:z0sring}
\zt_0(s) = \frac{2 (p-q) \rho_0^{-1}}{(N-1) (p+q)+1} s^{-2} + O(1/s\sqrt{s}),
\end{equation}
which after inverse Laplace transform and restoring $t=2 \tau /\mu_1$ back gives the following linear asymptotic growth of 
\begin{equation}
\label{eq:z0ring}
y_0(t) = z_0(t) = \frac{(p-q) \rho_0^{-1} \mu_1}{(N-1) (p+q) + 1} t + O(\sqrt{t}),~~~t \to \infty .
\end{equation}
% This means the average displacement of the tracer particle on a ring grows linearly with time for large $t$.

\item[] (b) Let us now look at the limits in the opposite order. We first take the thermodynamic limit and then we take $s \to 0$ limit. 
If $N$ is sent to infinity first, the sums in \eqref{eq:zls} become integrals. As a result the numerator of~\eqref{eq:zls} becomes
\begin{eqnarray}
\label{eq:intnum}
\frac{1}{N} \sum_{k=0}^{N-1} \frac{\ee^{\frac{2 \pi \ci k i}{N}}}{\lambda_k(s)}\Big{|}_{N \to \infty}=
\frac{1}{2}\int_{x=-1}^{1} \dd x~\frac{ \cos(\pi x i)}{s+4 \sin^2 (\pi x)} = \left(\frac{1}{2 \sqrt{s}} + O(1)\right)~e^{-i\sqrt{s}},
\end{eqnarray}
in $i \to \infty$ and $s \to 0$ limit while keeping $i\sqrt{s}$ finite.
Making use of the symmetry of the integrand under $x \rightarrow 1-x$, the denominator becomes
\begin{eqnarray}
\label{eq:intden}
1-\int_{x=0}^1 \frac{(2p-1) (\ee^{2 \pi \ci x}-1)+(2q-1) (\ee^{-2 \pi \ci x}-1)}{s+4 \sin^2(\pi x)} \dd x 
&=& -2 (p+q) \int_{x=0}^1 \frac{\ee^{2 \pi \ci x}-1}{s+4 \sin^2(\pi x)} \dd x + O(s^{1/2}) \\
&=& p+q + O(s^{1/2}). \nonumber 
\end{eqnarray}
Inserting the asymptotic forms from \eqref{eq:intnum} and \eqref{eq:intden} in \eqref{eq:zls}  we obtain $\zt_i(s)$ on infinite line for $s \to 0$ :
\begin{equation}
\label{eq:zlsline}
\zt_i(s) = \frac{p-q}{p+q} \rho_0^{-1} \frac{1}{s \sqrt{s}}~e^{-i\sqrt{s}} + O(s^{-1}),
\end{equation}
which after inverse Laplace transform and restoring $t=2 \tau /\mu_1$ gives
\begin{equation}
\label{eq:zlline}
y_i(t) = x_i(0) +z_0(t) \simeq  \rho_0^{-1}~\sqrt{2 \mu_1 t} \left[ \frac{i}{\sqrt{2 \mu_1 t}} + \frac{p-q}{p+q} \frac{1}{\sqrt{\pi}} 
\left \{ e^{-\frac{i^2}{2\mu_1t}} -\sqrt{\pi} \frac{|i|}{\sqrt{2 \mu_1 t}}~\Erfc\left( \frac{|i|}{\sqrt{2 \mu_1 t}}\right) \right \} \right],  
\end{equation}
where $\Erfc(z) = \frac{2}{\sqrt{\pi}} \int_{z}^\infty \ee^{-a^2} \dd a$. 
Hence the average displacement of the DTP ($i=0$) grows for large $t$ as  
\begin{equation}
\label{eq:z0line}
y_0(t) = z_0(t) \simeq  \frac{p-q}{p+q} \rho_0^{-1} \sqrt{\frac{2 \mu_1}{\pi}} \sqrt{t}.
\end{equation}
Similar late time growth $\sim \sqrt{t}$ for the average of the tracer position have been computed in the context of SSEP 
with a single driven tracer \cite{Burlatsky-92, Burlatsky-96}.
\end{itemize}
Comparing the large $t$ behaviors of $y_0(t)$ in equations \eqref{eq:z0ring} and \eqref{eq:z0line}, we see 
that the limits $N \rightarrow \infty$ and $t \rightarrow \infty$ do not commute. 
% At large times, the position of the tracer particle grows as $\sim t$ on a ring whereas it grows as $\sim \sqrt{t}$ on an infinite line.
When time $t$ is larger than the typical time required for an elementary transitions to occur but smaller than the time $ t\sim O( \rho_0^2 L^2)$ required 
for the particles to feel the finiteness of the ring, their positions grow as $\sim \sqrt{t}$. In this time scale the ring effectively acts as an infinite line.
On the other hand when $t \gg O( \rho_0^2 L^2)$, the finiteness of the ring comes into play and then their positions grow as $\sim t$.
We verify this behavior numerically in fig.~\ref{fig:yt}. Infact this crossover can entirely be described in terms of a nice crossover function which captures both 
the limits (a) and (b), discussed above. In the next we derive this crossover function.

\subsection{Finite size crossover}
\begin{figure}[!ht]
	\begin{center}
		\includegraphics[width=0.5\textwidth]{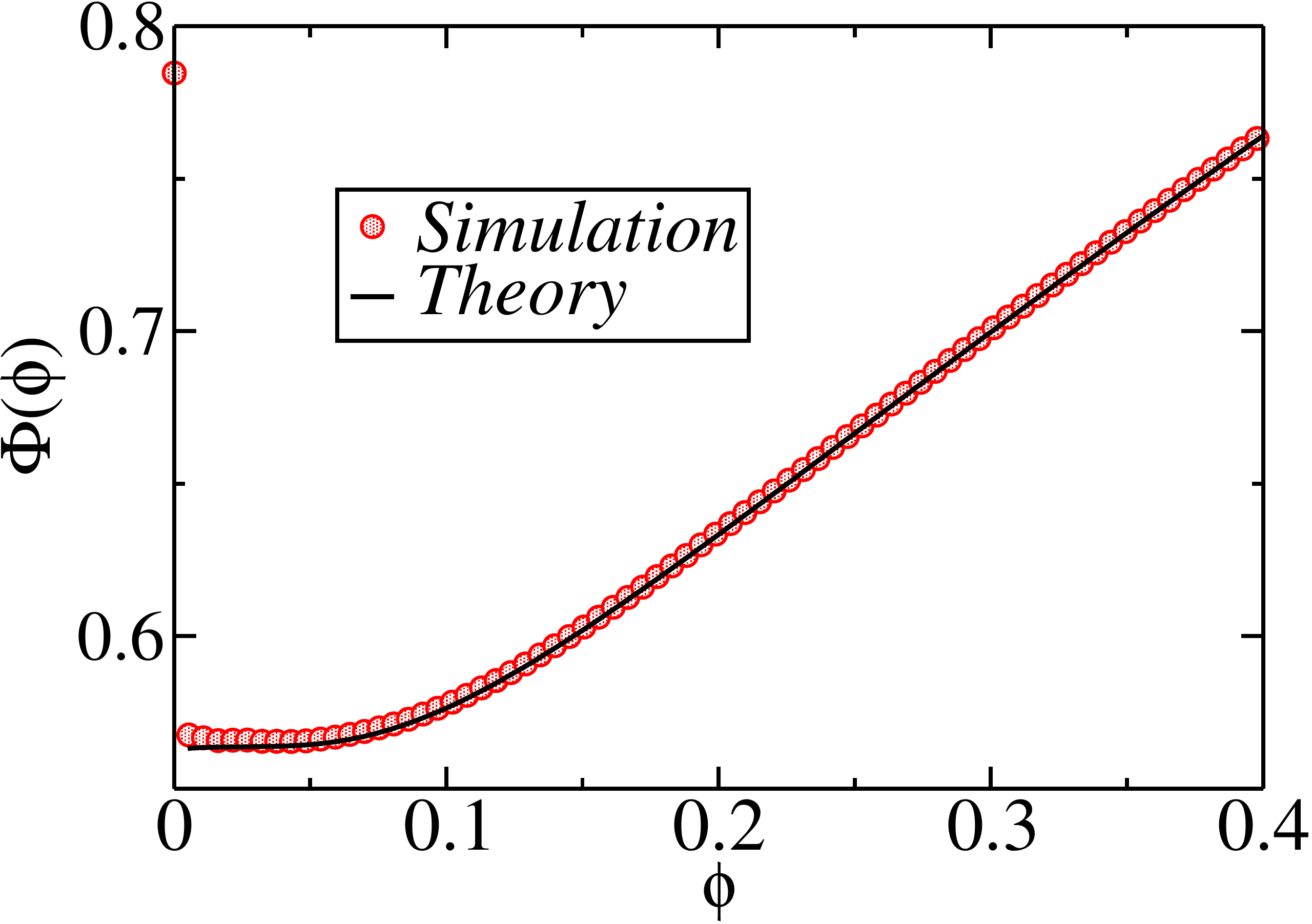} 
	\end{center}
	\caption{\small Numerical verification of the theoretical crossover function $\Phi(\phi)$ given in \eqref{eq:phi} for $L=1$, $N=200$, $p=1$ and $q=0$.}  
	\label{fig:crossover}
\end{figure}
% Better than computing the behavior of the DTP position in the ring and infinite line limits, we can obtain the function describing the crossover. 
\noindent
From the limiting cases~\eqref{eq:z0ring} 
and~\eqref{eq:z0line} we expect the crossover to be described by a function of the scaling variable $\phi = \frac{\tau}{N^2} = \frac{\mu_1 t}{2 N^2}$ ( or equivalently $\sigma = s N^2$ in the Laplace space) so that 
$\phi \to \infty$ and $\phi \to 0$ capture, respectively, the above two limits (a) and (b). To obtain the crossover function we start from the exact 
expression~\eqref{eq:zls} and compute $\tilde{z}_0(s)$ for the tracer particle
\begin{eqnarray}
\label{eq:crossoverlaplace}
\zt_0 \left(\frac{\sigma}{N^2}\right) &\sim_{N\rightarrow \infty}& \frac{2 (p-q) \rho_0^{-1}}{p+q} \frac{N^2}{\sigma} \frac{1}{N} \sum_{k=0}^{N-1} \frac{1}{\frac{\sigma}{N^2} + 4 \sin^2 \left( \frac{k \pi}{N}\right)} \nonumber \\
&\sim_{N\rightarrow \infty}& \frac{2 (p-q) \rho_0^{-1}}{(p+q)} \frac{N}{\sigma} \left[ \frac{N^2}{\sigma} + 2 \sum_{k=1}^{N/2} \frac{1}{\frac{\sigma}{N^2} + 4 \left( \frac{k \pi}{N}\right)^2} + 2 \sum_{k=1}^{N/2} 
\left( \frac{1}{\frac{\sigma}{N^2} + 4 \sin \left( \frac{k \pi}{N}\right)^2} - \frac{1}{\frac{\sigma}{N^2} + 4 \left( \frac{k \pi}{N}\right)^2} \right) \right] \\
&\sim_{N\rightarrow \infty}& \frac{2 (p-q) \rho_0^{-1}}{(p+q)} \frac{N}{\sigma} \left[ \frac{N^2}{\sigma} + 2 N^2 \sum_{k=1}^{\infty} \frac{1}{\sigma + 4 \pi^2 k^2} + \frac{N}{2 \pi} \int_{x=0}^{\pi/2}
\frac{x^2 -\sin^2(x)}{x^2 \sin^2(x)} \dd x \right] \nonumber \\
&\sim_{N\rightarrow \infty}& \frac{2 (p-q) \rho_0^{-1}}{(p+q)} N^3 \frac{1}{2 \sigma^{3/2}} \coth\left( \frac{\sqrt{\sigma}}{2}\right). \nonumber 
\end{eqnarray}
In the first line we took the limit of the denominator of~\eqref{eq:zls}, which does not introduce any complication. The sum on the numerator can then be taken care of by adding and subtracting the divergent part, 
as done on the second line. The first sum of the second line can now be evaluated exactly when $N \rightarrow \infty$, while the second sum converges to an integral which turns out to be subdominant. 
In the end the whole expression indeed converges to a function of the scaling variable $\sigma$.
The inverse Laplace transform is most easily performed on the crossover function expressed as a sum, \textit{i.e.} on the two first terms on the third line of Eq.\,\eqref{eq:crossoverlaplace}. In real space we get
\begin{equation}
\label{eq:crossoverreal}
z_0(t) \sim \frac{(p-q) \rho_0^{-1}}{p+q}\sqrt{2 \mu_1 t}~ \Phi\left(\frac{\mu_1 t}{2 N^2}\right),
\end{equation}
with the crossover function
\begin{equation}
\label{eq:phi}
\Phi(\phi) = \sqrt{\phi} + \frac{1}{2 \pi^2 \sqrt{\phi}} \sum_{k=1}^\infty \frac{1-\ee^{-4 \pi^2 k^2 \phi}}{k^2}.
\end{equation}
We can check that the asymptotic behaviors $\Phi(\phi) \sim_{\phi \rightarrow \infty} \sqrt{\phi}$ and $\Phi(\phi) \sim_{\phi \rightarrow 0} \pi^{-1/2}$ respectively give~\eqref{eq:z0ring} and~\eqref{eq:z0line} 
back in the limiting cases. The prediction~\eqref{eq:crossoverreal} is in very good agreement with the numerics, as shown in fig.\ref{fig:crossover}.

\subsection{Large $t$ scaling limit of $y_i(t)$ on infinite line}
\label{subsection:av-pos-sc}
\begin{figure}[!ht]
	\begin{center}
		\includegraphics[width=0.75\textwidth]{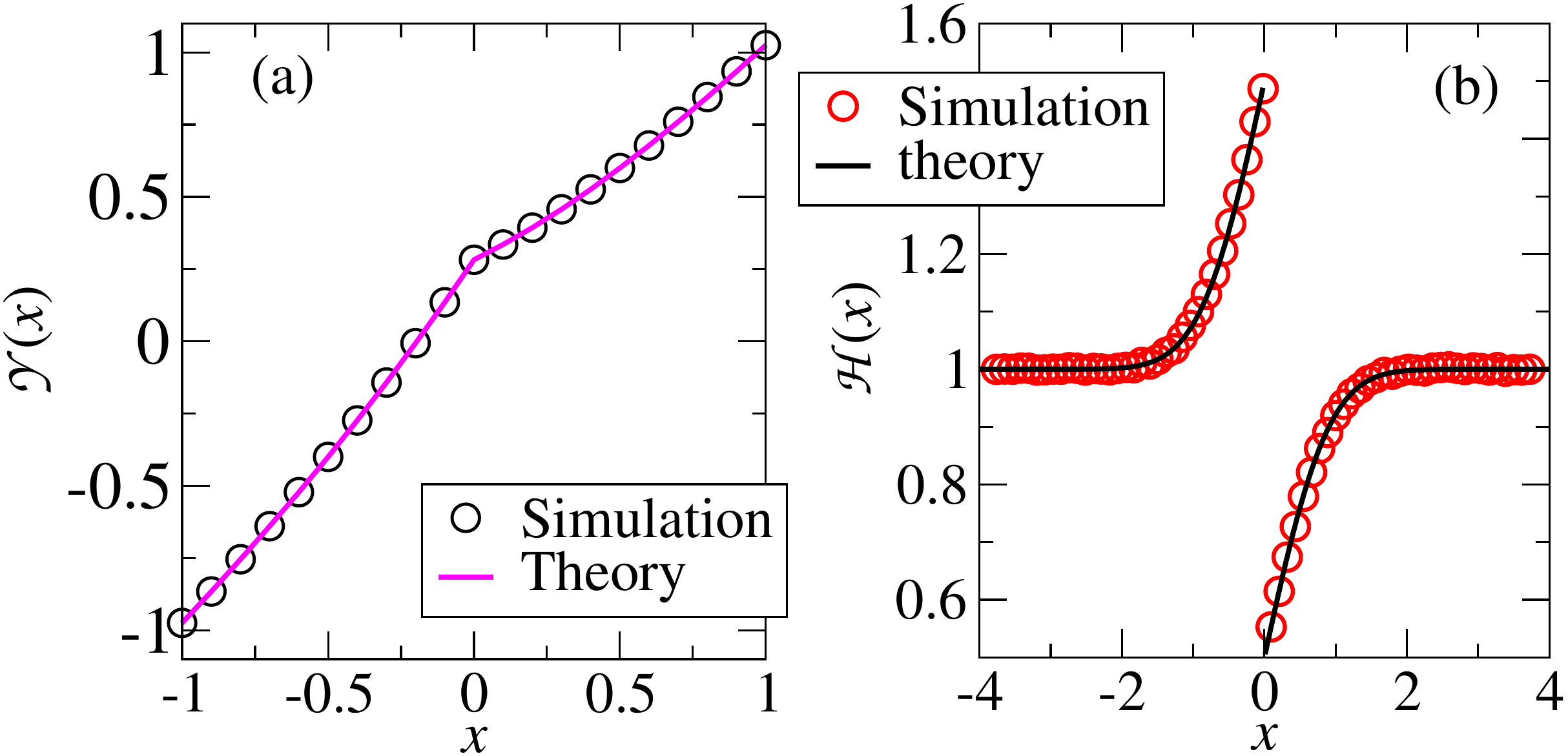} 
	\end{center}
	\caption{\small (a) Scaled average position profile $\cY(x)$ as a function of $x=\frac{i}{\sqrt{2\mu_1t}}$ for $t=700$. 
	                    The magenta solid line correxponds to the theoretical expression \eqref{eq:Ysol}.
	                (b) Scaled average gap profile $\cH(x)$ as a function of $x=\frac{i}{\sqrt{2\mu_1t}}$ for $t=700$. 
	                The black solid line corresponds to \eqref{eq:scalingH}.
                 The hopping rates of the DTP for this plot are $p=0.75$ and $q=0.25$.  
                 Jump distribution is uniform \emph{i.e.} $R(\eta)=1$.}  
	\label{fig:H}
\end{figure}

\noindent
Although in the previous section we have mainly looked at the growth of $y_0(t)$, but as a by product
we have also found the average position $y_i(t)$ of $i$-th particle 
in \eqref{eq:zlline} for case (b) where we take the thermodynamic limit before $t \to \infty$ limit.
Looking at \eqref{eq:zlline} we find that for large $t$, $y_i(t)$ has the following scaling form 
\begin{equation}
\label{eq:Yscaling}
y_i(t) = \rho_0^{-1} \sqrt{2 \mu_1 t}~ \cY \left( \frac{i}{\sqrt{2 \mu_1 t}} \right) + O(1),
\end{equation}
where the scaling function is given by 
\begin{equation}
\label{eq:Ysol}
\cY(x) = x + \frac{p-q}{p+q} \left[ \frac{\ee^{-x^2}}{\sqrt{\pi}} - |x|~ \Erfc(|x|)\right].
\end{equation}
This scaling function can also be computed in a different way as follows.
% \noindent              
The evolution equation for $y_i(t)$ on a line can be obtained either directly form the dynamics~\eqref{dyna-x} or from the 
equations on a ring by cutting the ring at $\frac{N}{2}$ and then sending $N$ to infinity. Both methods give
\begin{equation}
\label{eq:evoyi}
\dot{y}_i =\frac{\mu_1}{2} (y_{i+1} - 2 y_{i} + y_{i-1}) + \delta_{i,0} \frac{\mu_1}{2} \left( (2p-1) (y_1-y_0) + (2q-1) (y_{-1}-y_0)\right),
\end{equation}
where we recall that $\mu_1 = \int_{\eta=0}^1 \eta R(\eta) \dd \eta$ is the average of $\eta$. 
For large times, we look for solutions of this equation in the scaling form \eqref{eq:Yscaling}. Putting this form in \eqref{eq:evoyi} and taking $t \to \infty$ limit
we find that the function $\cY$ satisfies
\begin{equation}
\label{eq:Yode}
\cY''(x)+2x \cY'(x)-2 \cY(x) = \delta(x) \left[ (2 q-1) \cY'(0^-) - (2 p-1) \cY'(0^+)\right].
\end{equation}
The delta source at the origin implies that $\cY$ is continuous but its first derivative is discontinuous. 
By integrating both sides of \eqref{eq:Yode} over an infinitesimal segment from $0^-$ to $0^+$, one finds
\begin{equation}
\label{eq:Yp0}
p \cY'(0^+) = q \cY'(0^-).
\end{equation}
Two other boundary conditions are obtained by requiring that,
 particles far enough from the DTP are not perturbed. As a result average positions of the particles far from 
the DTP are equal to their initial positions. This implies 
\begin{equation}
\label{cY-x-lrg}
 \cY(x) \sim x~~\text{when}~~~|x| \rightarrow \infty.
\end{equation}
With the boundary conditions \eqref{eq:Yp0} and \eqref{cY-x-lrg}, one can easily solve \eqref{eq:Yode} to find $\cY(x)$ as given in \eqref{eq:Ysol}.
We observe nice agreement between this theoretical prediction and numerical measurements in fig.~\ref{fig:H}a.
\subsection{Average gap profile}
\label{subsection:av-gap-sc}
\noindent
From the average position profile $y_i(t)$ in \eqref{eq:Yscaling}, the average gap profile  $h_i(t) = \langle g_i(t) \rangle = y_{i+1}(t)-y_i(t)$ can be easily computed. 
Similar to $y_i(t)$, the average gap profile $h_i(t)$ also has a scaling form
\begin{equation}
\label{eq:gapscaling}
h_i(t) = \rho_0^{-1} ~\cH\left(\frac{i}{\sqrt{2 \mu_1 t}} \right) + O(t^{-1/2}),
\end{equation}
in the large $t$ limit. The scaling function $\cH(x)$ is obtained by taking the derivative of $\cY(x)$ in \eqref{eq:Yscaling}. We find 
\begin{equation}
\label{eq:scalingH}
\cH(x) = \cY'(x) = 1 - \frac{p-q}{p+q} ~\Sign(x)~ \Erfc\left(|x|\right),
\end{equation}
where $\Sign(x)=x/|x|$. 
In fig.~\ref{fig:H}b we compare this theoretical prediction with numerical measurements and the nice agreement between the two verifies our result.
%[Calculation, current]
\begin{figure}[!ht]
	\begin{center}
		\includegraphics[width=0.5\textwidth]{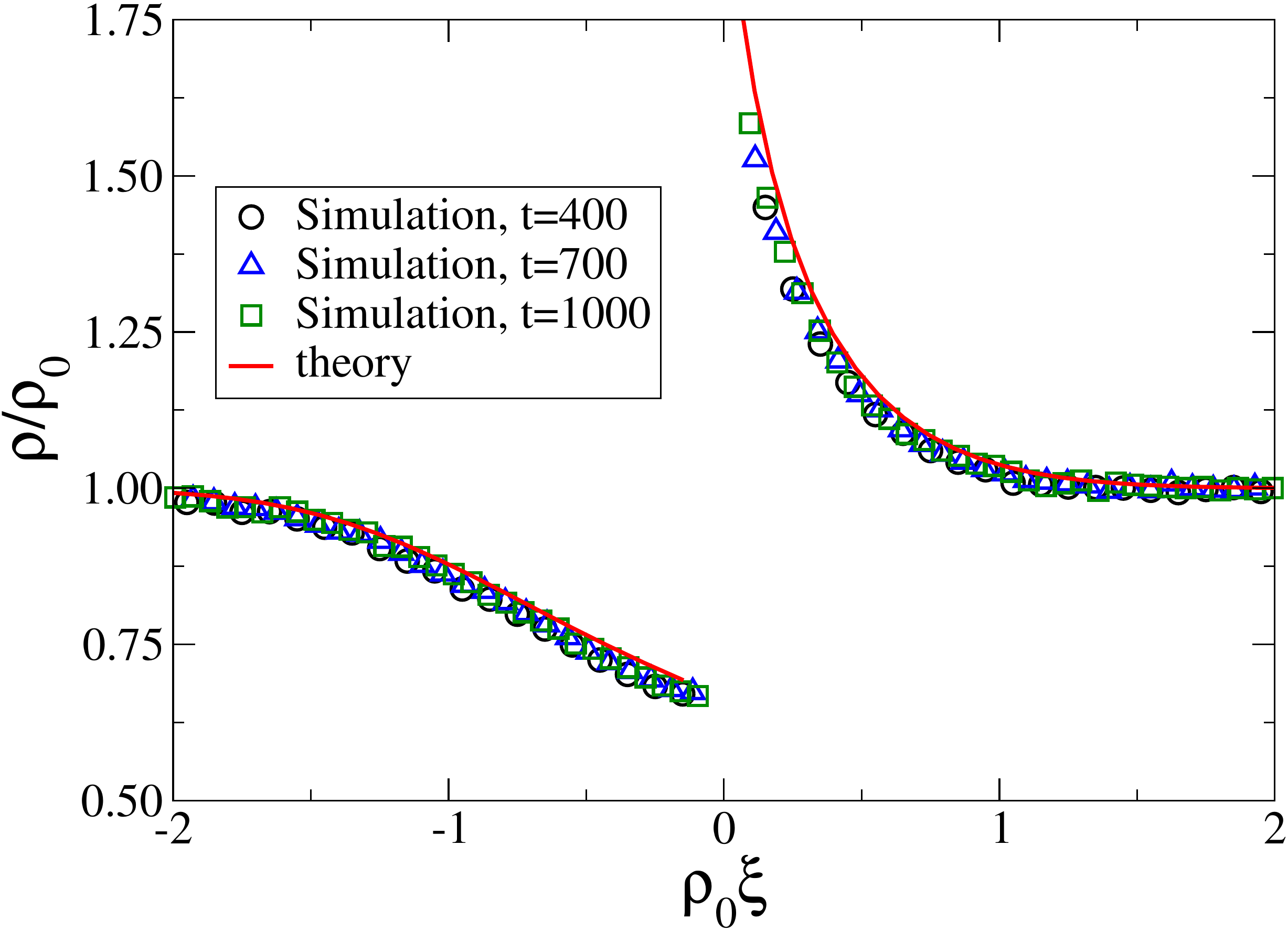} 
	\end{center}
	\caption{\small numerical verification of the particle density profile \eqref{den-scling} seen from the frame of the biased tracer particle. 
                  Parameters associated with this plot are : $p=0.75,~q=0.25$ and $\rho_0=200$. 
                  Jump distribution is uniform \emph{i.e.} $R(\eta)=1$. }  
	\label{fig:rho}
\end{figure}
\subsection{Particle density profile in the frame of the DTP}
\label{subsection:av-den-sc}
\noindent
From the knowledge of the average position profile $y_i(t)$ and the average gap profile $h_i(t)$, one can now find the mean particle density profile. 
In the frame of the DTP, the mean particle density at some space point $w$ at time $t$ is defined as 
\begin{equation}
 \rho(w,t) = \left \langle \sum_{i=-\infty}^\infty \delta[~w -(x_i(t)-x_0(t))~]\right\rangle, 
\end{equation}
where the angular average is taken over stochastic evolution. Since both $y_i(t)$ and $h_i(t)$ have scaling forms under the transformation 
\begin{equation}
 u=\frac{i}{\sqrt{2\mu_1 t}}, \label{u-def}
\end{equation}
for large $t$, we can expect that $\rho(w,t)$ also has a scaling behavior $\rho(w,t) \simeq \rho_0 \Omega\left(\frac{w}{\sqrt{2 \mu_1 t}} \right)$ for large $t$.
% \begin{equation}
%  \rho(z,t) \simeq \Omega(y),~~\text{where}~~~y=\frac{z}{\sqrt{2 \mu_1 t}}~. 
% \end{equation}
One can observe this scaling behavior in numerical simulations. The question now is : what is the expression of $\Omega(\xi)$ ? 
To find that, let us start with the discrete picture. In terms of the average gaps $h_i(t)$s, the average position of $i$-th particle with respect to the DTP is given by
\begin{equation}
 \tilde{y}_i(t)=y_i(t)-y_0(t) = \sum_{l=0}^{i-1}h_l(t), \label{pos-gap-rela}
\end{equation}
which in large $t$ limit becomes 
\begin{equation}
 \xi(u) = \rho_0^{-1} \int_0^u ~\cH(a)~\dd a,  \label{pos-indx-rela}
\end{equation}
where $u$ is given in \eqref{u-def} and $\xi(u)=\frac{\tilde{y}_i}{\sqrt{2 \mu_1 t}}$. 
On the other hand, the average density near the $i$-th particle at time $t$ can approximately be given by $\rho(\tilde{y}_i,t) \simeq \frac{2}{h_{i-1}(t)+h_i(t)},$
which in large $t$ limit gives
\begin{equation}
 \rho(\xi(u)) = \frac{\rho_0}{\cH(u)} + O \left(\frac{1}{\sqrt{t}} \right), \label{den-scling}
% ~~~\text{where},~~ \xi=\frac{y_i}{\sqrt{2 \mu_1 t}}. 
\end{equation}
Equations \eqref{pos-indx-rela} and \eqref{den-scling} together constitute the density profile in parametric form. 
We compare the theoretical expression~\eqref{pos-indx-rela}-\eqref{den-scling} of $\rho$ as a function of $\xi$ 
with numerical measurements in fig.~\ref{fig:rho} and find quite good agreement.  
We observe that the average density profile gets modulated because of the biased motion of the tracer particle; the system is denser 
in front of the biased TP and sparser at the back of it.
%This density modulation $\rho - \rho_0$ decays as one moves away from the biased TP and decays very fast as 
%$\sim \frac{\ee^{-\rho_0^2 \xi^2}}{\xi}$ when $|\xi| \rightarrow \infty$ 
%\emph{i.e.} on both sides of the biased tracer. 
This density modulation $\rho - \rho_0$ decays very fast as one moves away from the tracer on both sides, as $\sim \frac{\ee^{-\rho_0^2 \xi^2}}{\xi}$ when $|\xi| \rightarrow \infty$.

A similar phenomenon is observed when a biased tracer is present in a one-dimensional simple exclusion process \cite{Burlatsky-92,Burlatsky-96}. 
As in our case, the velocity of the tracer decays as $\sim 1/\sqrt{t}$~\cite{Burlatsky-92}. The length scale over which one observes 
the effect of the tracer also scales as $\sqrt{t}$. Moreover we note that the decay of the density perturbation at large 
distances from the DTP is exactly the same as in the RAP case \emph{i.e.} $\sim \frac{\ee^{-\rho_0^2 \xi^2}}{\xi}$ for $|\xi| \rightarrow \infty$ 
(see equations~(26) and~(31) of \cite{Burlatsky-96}). In contrast, the phenomenon is different for driven tracers in SSEP of higher dimensions, 
where the velocity of the tracer is finite, the density around the tracer reaches a stationary profile without going to a scaling limit and the decay of the density 
modulation is exponential everywhere except at the back of the driven tracer, where it becomes algebraic~\cite{Coninck-97,Benichou-01,Benichou-00}.

\section{Correlations}
\label{section:2pt}
\begin{figure}[!ht]
	\begin{center}
		\includegraphics[width=0.5\textwidth]{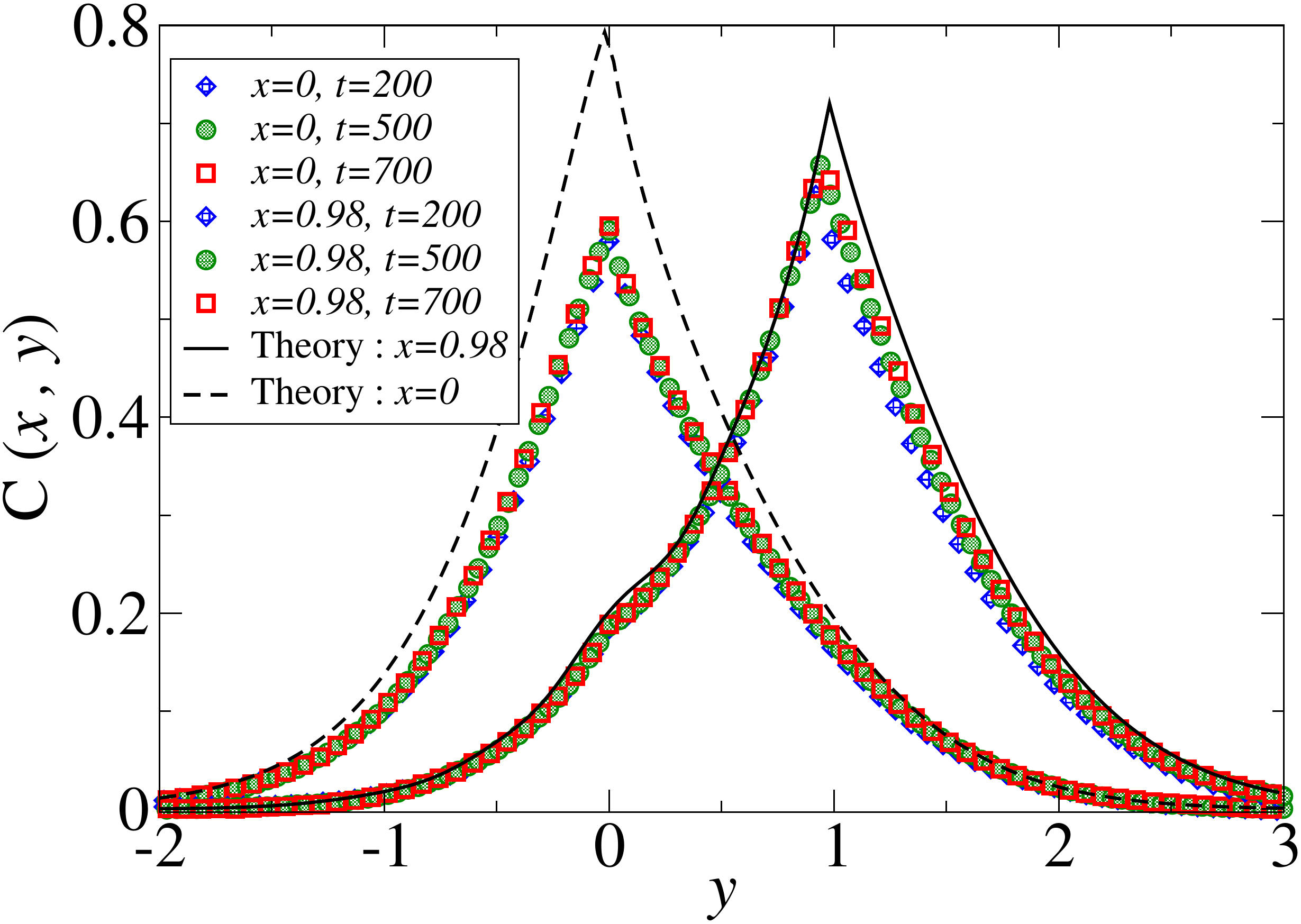} 
	\end{center}
	\caption{\small Numerical verification of the scaling form \eqref{eq:scalingc} for $c_{i,j}(t)$. 
                        Solid black lines correspond to approximate solution for $\cC(x,y) \simeq \cC_{ub}(x,y)+\frac{p-q}{2}~\cC_1(x,y)$ obtained 
                        in sec. \ref{pertur-sol-C}. The function $\cC_{ub}(x,y)$ is given in \eqref{c-inf} and $\cC_1(x,y)$, given in \eqref{eq:C1psi}, is 
                        evaluated using \eqref{eq:exppsi}. We performed the sum over $m$ in \eqref{eq:exppsi} until $m=8$. 
                        For $x=0$ the theoretical curve does not match with the numerical results because we have only 
                        considered first order perturbation theory. Indeed, by symmetry $\cC_1(x,y)$ vanishes at $x=y=0$, 
                        so that the first dominant correction comes from higher orders in $\epsilon$. 
                        The parameters associated with this plot are $p=0.75,~q=0.25$ and $N=200$. 
                        Jump distribution is uniform \emph{i.e.} $R(\eta)=1$. }  
	\label{fig:Csc}
\end{figure}
\noindent
In this section we study the two-point connected correlation function of the positions,
\begin{equation}
\label{eq:defcij}
c_{i,j}(t) = \langle x_i(t) x_j(t) \rangle -y_i(t) y_j(t),
\end{equation}
and of the gaps,
\begin{equation}
\label{eq:defdij}
d_{i,j}(t) = \langle g_i(t) g_j(t) \rangle -h_i(t) h_j(t).
\end{equation}
In the previous section we have seen that both the mean position profile $y_i(t)$ and the mean gap profile $h_i(t)$, have scaling forms when index $i$ is scaled 
by $\sqrt{2\mu_1 t}$. When $p=q=1/2$, \emph{i.e.} when all the particles are moving symmetrically, the position correlation function $c_{i,j}(t)$ 
have been computed by Rajesh and Majumdar \cite{Rajesh-01}. Looking at their result \eqref{C_r-RM} with $\alpha=\beta=1/2$,  we find that $c_{i,j}(t)$ has 
a scaling form  as a function of the scaling variable $u=\frac{|i-j|}{\sqrt{2\mu_1 t}}$.
On the basis of these facts we expect that both the correlation functions $c_{i,j}(t)$ and $d_{i,j}(t)$ have well defined scaling limits under the transformations 
$i \to x= \frac{i}{\sqrt{2\mu_1 t}}$ and $j \to y= \frac{j}{\sqrt{2\mu_1 t}}$ for large $t$. To support this hypothesis, let us first present our numerical results.

% \subsection{Numerical results}
% \noindent
% We have numerically simulated the RAP dynamics with one driven TP. 
We have numerically measured the pair position correlations and pair gap correlations defined, respectively, in \eqref{eq:defcij} and \eqref{eq:defdij} as a function of $j$ 
for different fixed values of $i$ and $t$. In fig.~\ref{fig:Csc} we plot $\frac{\rho_0^2c_{i,j}(t)}{\sqrt{2\mu_1 t}}$ as a function of $y=\frac{j}{\sqrt{2 \mu_1 t}}$ 
for  $x=0$ and $0.98$, and for three different values of $t=200,~500$ and $700$ and we observe a clear and excellent data collapse. 
This verifies our hypothesis and implies the following scaling form of $c_{i,j}(t)$ for large $t$ :
\begin{equation}
\label{eq:scalingc}
c_{i,j}(t) = \rho_0^{-2} \sqrt{2 \mu_1 t}~\cC \left( \frac{i}{\sqrt{2 \mu_1 t}}, \frac{j}{\sqrt{2 \mu_1 t}}\right) + O(1).
\end{equation}

\noindent
Similar to $c_{i,j}(t)$ the gap correlation function $d_{i,j}(t)$ also has a scaling form (numerically verified but not presented here)
\begin{equation}
\label{eq:scalingd}
d_{i,j}(t) = \frac{\rho_0^{-2}}{\sqrt{2 \mu_1 t}} ~\cD \left( \frac{i}{\sqrt{2 \mu_1 t}}, \frac{j}{\sqrt{2 \mu_1 t}}\right) + O(t^{-1}), \qquad i \neq j,
\end{equation}
where the scaling function $\cD(x,y)$ is related to $\cC(x,y)$ as
\begin{equation}
\label{eq:linkcd}
\cD(x,y) = \partial_x \partial_y \cC(x,y).
\end{equation}
However this scaling form \eqref{eq:scalingd} is valid only for off-diagonal gap correlation functions. For diagonal gap correlations we in fact observe numerically 
(see fig. \ref{fig:V}) that, $d_{i,i}(t)$ is of order one not of order $1/\sqrt{t}$. Hence for $i=j$ line, we consider the following scaling form for $d_{i,j}(t)$ :
\begin{equation}
\label{eq:scalingdd}
d_{i,i} = \rho_0^{-2} \cV \left( \frac{i}{\sqrt{2 \mu_1 t}}\right) + \frac{\rho_0^{-2}}{\sqrt{2 \mu_1 t}} \left[ \cV_1 \left( \frac{i}{\sqrt{2 \mu_1 t}}\right)  
+ \cD \left( \frac{i}{\sqrt{2 \mu_1 t}}, \frac{i}{\sqrt{2 \mu_1 t}}\right) \right] + O(t^{-1}).
\end{equation}
Equations~\eqref{eq:scalingd} and~\eqref{eq:scalingdd} are supported by numerical evidences.
Our next aim is to compute these scaling functions $\cC(x,y)$, $\cD(x,y)$, $\cV(x)$ and $\cV_1(x)$ analytically.

%[Numerics first to justify scaling]
\subsection{Computation of $\cC(x,y)$}
\label{C-diff-eq}
\noindent
To compute $\cC(x,y)$ we start with the discrete evolution equation for $c_{i,j}(t)$ which 
can be obtained from the dynamics of the positions~\eqref{dyna-x}. It reads as 
\begin{eqnarray}
\label{eq:evocij}
\dot{c}_{i,j} &=& \frac{\mu_1}{2} (c_{i+1,j} + c_{i-1,j} + c_{i,j+1} + c_{i,j-1} - 4 c_{i,j}) \nonumber \\ 
&&+ \delta_{i,j} \frac{\mu_2}{2} (c_{i+1,i+1} - 2 c_{i+1,i} + 2 c_{i,i} - 2 c_{i-1,i} + c_{i-1,i-1} + (y_{i+1}-y_i)^2 + (y_{i-1}-y_i)^2) \\
&&+ \delta_{i,0} \frac{\mu_1}{2} ((2p-1) (c_{1,j}-c_{0,j}) + (2q-1) (c_{-1,j}-c_{0,j})) 
+ \delta_{j,0} \frac{\mu_1}{2} ((2p-1) (c_{i,1}-c_{i,0}) + (2q-1) (c_{i,-1}-c_{i,0})) \nonumber \\ 
&&+ \delta_{i,0} \delta_{j,0} \frac{\mu_2}{2} ((2p-1) (c_{1,1} - 2 c_{0,1} + c_{0,0} + (y_1-y_0)^2) + (2q-1)(c_{-1,-1} - 2 c_{0,-1} + c_{0,0} + (y_{-1}-y_0)^2)), \nonumber
\end{eqnarray}
where $y_i(t)=\langle x_i(t)\rangle$.
We are interested in finding the solution of this equation in the form \eqref{eq:scalingc} for large $t$. 
For this, one can follow the Fourier-Laplace transform method as used in solving \eqref{eq:evoziring}, to show 
that $c_{i,j}(t)$ indeed has the scaling form \eqref{eq:scalingc} for large $t$. But performing such analysis 
involves two coupled integral equations arising from the self consistency conditions and that makes it hard to solve.
Instead assuming  $c_{i,j}(t)$ has the scaling form \eqref{eq:scalingc} for large $t$, we insert this scaling form in 
the discrete equations \eqref{eq:evocij} and take large $t$ limit to obtain the following differential equation for $\cC(x,y)$
% Inserting the scaling form of $c_{i,j}(t)$ 
% from \eqref{eq:scalingc} in \eqref{eq:evocij} first and then taking large $t$ limit, we obtain 
\begin{eqnarray}
\label{eq:evoCscaling}
[\partial_x^2 + \partial_y^2 +2 x \partial_x + 2 y \partial_y -2] \cC(x,y) 
&=& \delta(x-y) \frac{2 \mu_2}{\mu_1} [\partial_y \cC|_{x-y=0^-}-\partial_x \cC|_{x-y=0^-} - \cH(x)^2] \nonumber \\ 
&&+ \delta(x) [(2q-1) \partial_x \cC|_{x=0^-}-(2p-1) \partial_x \cC|_{x=0^+}] \\ &&+ \delta(y) [(2q-1) \partial_y \cC|_{y=0^-}-(2p-1) \partial_y \cC|_{y=0^+}], \nonumber 
\end{eqnarray}
in the leading order. Here $\cH(x)$ is the average gap profile. We solve this equation for $\cC(x,y)$ and verify the solution with numerical measurements.

At first glance, equation~\eqref{eq:evoCscaling} seems complicated because of the self-consistent terms on the right hand side (RHS).
However we can simplify it further.
We start with the $\delta(x)$ and $\delta(y)$ terms on RHS. From numerical measurements we have seen that $\cC(x,y)$ is continuous across $x=0$ 
but its derivative is possibly 
discontinuous. Integrating both sides of \eqref{eq:evoCscaling} from $x=0^-$ to $x=0^+$, 
we find that the derivative should satisfy $q \partial_x \cC(0^-,y) = p \partial_x \cC(0^+,y)$ for all $y$. 
A symmetric argument can be applied to the $y=0$ line too. Since the equation is of the second order, we expect  that the knowledge of two 
matching conditions at each non-analiticity is enough to determine the solution. In summary, the $\delta(x)$ and $\delta(y)$ terms on RHS of \eqref{eq:evoCscaling} can 
equivalently be replaced by imposing the boundary conditions
\begin{eqnarray}
\label{eq:evoCscalingbc}
\cC(0^-,y) = \cC(0^+,y), &\qquad& q \partial_x \cC(0^-,y) = p \partial_x \cC(0^+,y), \\
\cC(x,0^-) = \cC(x,0^+), &\qquad& q \partial_y \cC(x,0^-) = p \partial_y \cC(x,0^+). \nonumber
\end{eqnarray}
\begin{figure}[!ht]
	\begin{center}
		\includegraphics[width=0.5\textwidth]{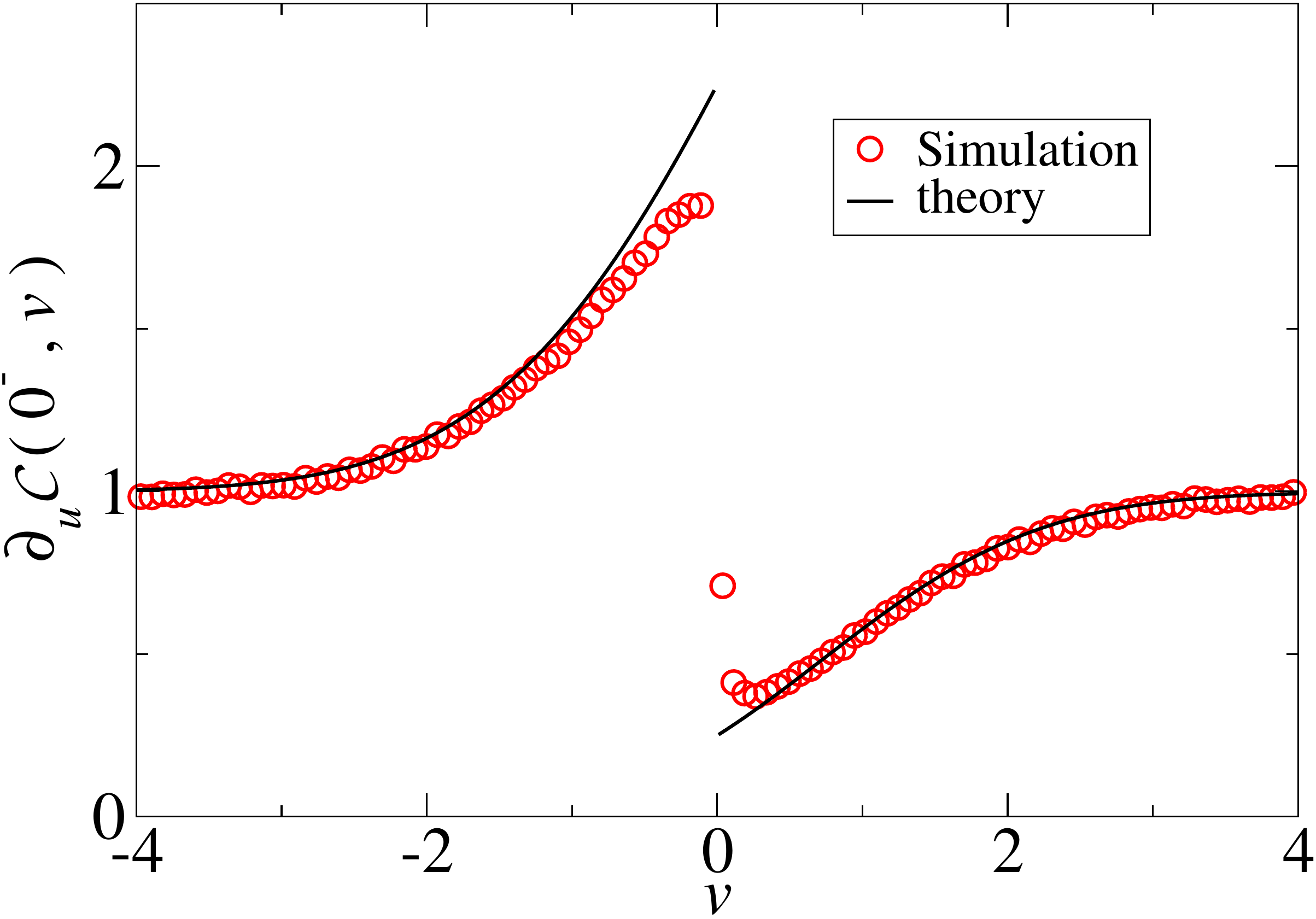} 
	\end{center}
	\caption{\small Numerical verification of \eqref{eq:dCudiag} as a function of $v$ for $N=200$, $t=700$, $p=0.75$ and $q=0.25$. 
                Circles represent $\partial_u \cC(0^-,v)$ obtained from numerical measurements and the solid line 
                is $\frac{\mu_2}{2 (\mu_1 - \mu_2)} \cH\left( \frac{v}{2}\right)^2$. 
                Jump distribution is uniform \emph{i.e.} $R(\eta)=1$.}  
	\label{fig:ducdiag}
\end{figure}

\noindent
Let us now simplify the $\delta(x-y)$ term and for that we follow the same procedure as done for the $\delta(x)$ term, 
namely integrate both sides of \eqref{eq:evoCscaling} across $x=y$ line. We therefore make the coordinate transformation $u=x-y$ and $v=x+y$. 
In terms of the transformed variables $u$ and $v$, Eq.\,\eqref{eq:evoCscaling} reads
\begin{equation}
\label{eq:evoCscalinguv}
[\partial_u^2 + \partial_v^2 +u \partial_u + v \partial_v -1] \cC(u,v) = - \delta(u) \frac{\mu_2}{\mu_1} \left[2 \partial_u \cC(0^-,v) 
+ \cH\left( \frac{v}{2}\right)^2 \right],
\end{equation}
where the $\delta$-source terms for $x=0$ and $y=0$ are replaced by the boundary conditions~\eqref{eq:evoCscalingbc}. We integrate again across $u=0$ line from below to above.
As evidenced from numerical measurements of $c_{i,j}(t)$ in fig.~\ref{fig:Csc}, $\cC$ is continuous at $u=0$. We therefore get an equation for the discontinuity 
of the first derivative across $u=0$ line,
\begin{equation}
\label{eq:discdCu}
\mu_1 (\partial_u \cC(0^+,v) - \partial_u \cC(0^-,v) ) = -2 \mu_2 \partial_u \cC(0^-,v) - \mu_2 \cH\left( \frac{v}{2}\right)^2.
\end{equation}
Using the symmetry of $\cC$ under reflection with respect to the diagonal, we get 
\begin{equation}
\label{eq:dCudiag}
\partial_u \cC(0^-,v) = - \partial_u \cC(0^+,v) = \frac{\mu_2}{2 (\mu_1 - \mu_2)} \cH\left( \frac{v}{2}\right)^2,
\end{equation}
which is verified numerically in fig.~\ref{fig:ducdiag}. Now, inserting the result~\eqref{eq:dCudiag} in the RHS of \eqref{eq:evoCscaling} 
and transforming back to original $(x,y)$ coordinates, we get
\begin{equation}
\label{eq:evoCscalingsimp}
[\partial_x^2 + \partial_y^2 +2 x \partial_x + 2 y \partial_y -2] \cC(x,y) = - \delta(x-y) \frac{2 \mu_2}{\mu_1-\mu_2} \cH(x)^2,
\end{equation}
with boundary conditions \eqref{eq:evoCscalingbc}. Other boundary conditions come from the fact that for large $x$ and $y$, \emph{i.e} when both 
the particles are far from the driven tracer, the correlation among their positions should be equal to the correlation 
function of the non-driven system, given by Eq.\,\eqref{C_r-RM} with $\alpha=\beta=1/2$. 
% This means for $|x| \to \infty$ and $|y| \to \infty$, 
This means
\begin{eqnarray}
&&
\cC(x,y) \simeq \cC_{ub}(x,y) = \frac{\mu_2}{\sqrt{2 \pi} (\mu_1 - \mu_2)}~g(x-y),~~~
\text{for}~~|x| \to \infty,~~~|y| \to \infty,
% \begin{cases}
% &|x| \to \infty \\
% % &\text{and/or}\\
% &|y| \to \infty\\ 
% \end{cases},
\label{c-inf}\\
&&~~~~~~~~~~~\text{where},
~~g(u)=e^{-\frac{u^2}{2}} - \sqrt{\frac{\pi}{2}} ~|u|~\Erfc\left( \frac{|u|}{\sqrt{2}}\right),  \label{g_u}
\end{eqnarray}
and the subscript ``ub'' denotes unbiased case. 
We now have to solve the differential equation \eqref{eq:evoCscalingsimp} with boundary conditions \eqref{eq:evoCscalingbc} and \eqref{c-inf}. 
Computing the full solution for $\cC(x,y)$ for arbitrary $p$ and $q$ in a closed form seems difficult. We are however able to solve \eqref{eq:evoCscalingsimp} 
perturbatively by expanding $\cC(x,y)$ in powers of the drive strength $\epsilon=p-q$.
% [small p-q? result far from the tracer, more general result? q=0?]

\begin{figure}[!ht]
	\begin{center}
		\includegraphics[width=0.5\textwidth]{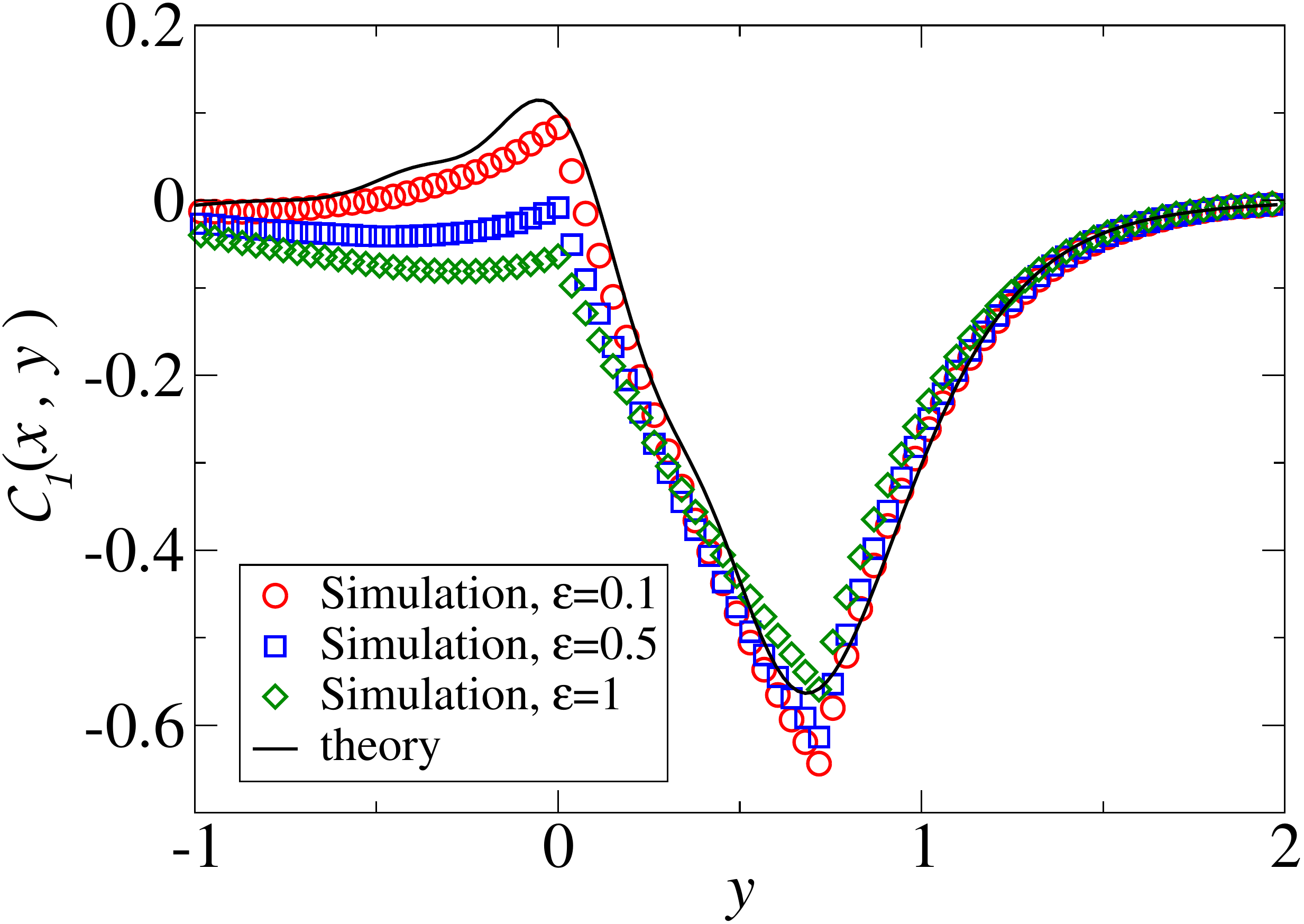} 
	\end{center}
	\caption{\small Comparison of theoretically obtained $\cC_1(x,y)$ with numerical measurements (circles) for $x=0.756$. 
                        The wavy and rounded nature of the theoretical solution is due to restricting the sum in \eqref{eq:exppsi} up to $m=8$ while evaluating in Mathematica.
                        The simulation data (symbols) are obtained by subtracting $d_{i,j}(t)$ with $\epsilon=0$ from $d_{i,j}(t)$ with $\epsilon \neq 0$.
                        The parameters associated with this plot are : $N=200$, $t=700$ and $p+q=1$. Jump distribution is uniform \emph{i.e.} $R(\eta)=1$.}
	\label{fig:C1}
\end{figure}

\subsubsection{Perturbative expansion in $\epsilon = p-q$}
\label{pertur-sol-C}
\noindent
Let us consider the following expansions of the functions $\cH$ and $\cC$ in powers of $\epsilon$,
\begin{eqnarray}
\label{eq:expansionHC}
\cH(x) &=& 1 - \frac{\epsilon}{p+q}~ \Sign(x)~ \Erfc(|x|), \\
\cC(x,y) &=& \cC_0(x,y) + \frac{\epsilon}{2} ~\cC_1(x,y) + \frac{\epsilon^2}{4} ~\cC_2 (x,y) + \ldots. \nonumber
\end{eqnarray}
% where at the zeroth order we have the correlation function $\cC_0(x,y)$ of the un-driven system and the $\cC_i(x,y)$ are order $O(\epsilon^i)$ corrections to $\cC_0(x,y)$. 
This expansion of $\cC(x,y)$ provides a systematic way of solving equation~\eqref{eq:evoCscalingsimp} order by order in $\epsilon$. 
Indeed, inserting the expansions~\eqref{eq:expansionHC} in the evolution equation~\eqref{eq:evoCscalingsimp}, we get equations for each $\cC_i(x,y)$ with previous order functions 
$\cC_j(x,y),~j<i$ appearing as source. 
In this paper we compute $\cC(x,y)$ till first order. However our method can be generalized to obtain higher order solutions.
At order $\epsilon^0$ we have 
\begin{equation}
\label{eq:evoC0scaling}
[\partial_x^2 + \partial_y^2 +2 x \partial_x + 2 y \partial_y -2] \cC_0(x,y) = \delta(x-y) \frac{8 \mu_2}{\mu_1-\mu_2}.
\end{equation}
As this equation physically corresponds to system without drive, we have $\cC_0(x,y) \equiv \cC_{ub}(x,y)$ where $\cC_{ub}$ is given in \eqref{c-inf}.

Let us now focus at order $\epsilon$. If we choose to keep the $\delta(x)$ and $\delta(y)$ source terms of 
equation~\eqref{eq:evoCscaling} instead of taking them as boundary conditions, then using \eqref{eq:dCudiag} and \eqref{eq:expansionHC} we get
% and the expansion \eqref{eq:expansionHC} we get 
\begin{equation}
\label{eq:evoC1scaling}
[\partial_x^2 + \partial_y^2 +2 x \partial_x + 2 y \partial_y -2] \cC_1(x,y) = \delta(x-y) \frac{8 \mu_2}{\mu_1-\mu_2} \frac{\Sign(x) \Erfc(|x|)}{p+q}
- 4 \delta(x)\partial_x \cC_{ub}(0,y) - 4 \delta(y) \partial_y \cC_{ub}(x,0).
\end{equation}
% where we have used explicit expansion of $\cH(x)$ from \eqref{eq:expansionHC}. 
Explicit expression of $\partial_x \cC_{ub}(0,y)$ can be obtained from \eqref{c-inf} as 
\begin{equation}
\label{eq:dC0}
\partial_x \cC_{ub}(0,y) = \frac{\mu_2}{2 (\mu_1-\mu_2)} ~ \Sign(y) ~\Erfc \left( |y|\right).
\end{equation}
Similarly $\partial_y \cC_{ub}(x,0)$ can also be obtained. Going to (tilted) polar coordinates 
$(x,y) = (r \cos \left( \theta + \frac{\pi}{4} \right), r \sin \left( \theta + \frac{\pi}{4} \right))$ and 
using~\eqref{eq:dC0}, we rewrite equation \eqref{eq:evoC1scaling} as
\begingroup\makeatletter\def\f@size{9}\check@mathfonts
\def\maketag@@@#1{\hbox{\m@th\large\normalfont#1}}%
\begin{align}
\label{eq:evoC1scalingpol}
&&\left[\partial_r^2 +  \frac{1}{r} \partial_r + \frac{1}{r^2} \partial_\theta^2 +2 r \partial_r -2 \right] \cC_1(r,\theta) 
= \frac{2 \mu_2}{\mu_1-\mu_2} \frac{\Erfc\left(r/\sqrt{2}\right)}{r} \bigg[ &\frac{2 \sqrt{2}}{p+q} \left(\delta\left(\theta \right)
-\delta \left( \theta-\pi \right)\right)  \\
&&-& \delta \left(\theta -\frac{\pi}{4}\right)+\delta \left(\theta+\frac{3 \pi}{4}\right) 
- \delta\left(\theta +\frac{\pi}{4}\right)+\delta \left(\theta-\frac{3 \pi}{4}\right) \bigg].\nonumber
\end{align}\endgroup
% \begin{eqnarray}
% \label{eq:evoC1scalingpol}
% \left[\partial_r^2 +  \frac{1}{r} \partial_r + \frac{1}{r^2} \partial_\theta^2 +2 r \partial_r -2 \right] \cC_1(r,\theta) 
% &=& \frac{2 \mu_2}{\mu_1-\mu_2} \frac{\Erfc\left(r/\sqrt{2}\right)}{r} \bigg[ \frac{2 \sqrt{2}}{p+q} \left(\delta\left(\theta \right)-\delta \left( \theta-\pi \right)\right) \nonumber \\
% &&- \delta \left(\theta -\frac{\pi}{4}\right)+\delta \left(\theta+\frac{3 \pi}{4}\right) 
% - \delta\left(\theta +\frac{\pi}{4}\right)+\delta \left(\theta-\frac{3 \pi}{4}\right) \bigg].
% \end{eqnarray}
The boundary conditions for the above equation are : $\cC_1(r,\theta)|_{r\to0}$ is finite and $\cC_1(r,\theta)|_{r\to \infty}\to 0$. 
We observe that $\cC_1(r,\theta)$ can be written as 
\begin{equation}
\label{eq:C1psi}
\cC_1(r,\theta) = \frac{2 \mu_2}{\mu_1-\mu_2} \left( \frac{2 \sqrt{2}}{p+q}~ \psi(r,\theta) - \psi\left(r,\theta + \frac{\pi}{4}\right) 
- \psi\left(r,\theta - \frac{\pi}{4}\right) \right),
\end{equation}
where $\psi(r,\theta)$ satisfies 
\begin{equation}
\label{eq:psi}
\left[\partial_r^2 +  \frac{1}{r} \partial_r + \frac{1}{r^2} \partial_\theta^2 +2 r \partial_r -2 \right] \psi(r,\theta) 
=  \left(\delta\left(\theta \right)-\delta \left( \theta-\pi \right)\right) \frac{\Erfc(r/\sqrt{2})}{r},
\end{equation}
This equation can be solved by expanding both $\delta(\theta)$ and $\psi(r,\theta)$ as 
\begin{equation}
\label{eq:exppsi}
\delta(\theta) = \frac{1}{2 \pi} \sum_{l=-\infty}^\infty \ee^{\ci l \theta},
~\text{and}~\psi(r,\theta) = \sum_{m=0}^\infty 2 \cos ((2m+1)\theta)~ \psi_m(r), 
\end{equation}
for $l,m$ Integers. 
% since the even $l$ terms cancel out on the RHS of~\eqref{eq:psi}.
Inserting this expansion in \eqref{eq:psi}, one obtains a radial differential equation for each $m$, which one has to solve with boundary conditions :
\begin{equation}
\psi_0(r) |_{r \to 0} = \text{finite},~~\psi_{m>0}(r) |_{r \to 0} =0, \text{~~and~~~}  \psi_m(r) |_{r \to \infty} = 0~~~\forall m.
\end{equation}
Since the calculation of $\psi_m(r)$ is long and technical, we 
do not present it in the main body of the paper but rather present it in appendix~\ref{section:appC1}. 
Inserting $\psi_m(r)$ (from appendix~\ref{section:appC1}) in \eqref{eq:exppsi} we obtain $\psi(r,\theta)$, using which in \eqref{eq:C1psi} we evaluate $\cC_1(r,\theta)$.
In fig.~\ref{fig:C1} we compare the theoretically computed $\cC_1(r,\theta)$ with the numerical measurements and observe nice agreement.

As it seems the explicit expression of $\psi(r,\theta)$ (see appendix~\ref{section:appC1}) is not very illuminating, here we look at the asymptotic behavior of $\psi(r,\theta)$ for 
small and large $r$ values. We find that close to $r = 0$ the function $\psi(r,\theta)$ behaves as (see appendix~\ref{sml_lrg_r-psi} for details)
\begin{equation}
\label{eq:psismallrf}
% \psi(r,\theta) = \frac{r \log r \cos \theta}{\pi} +\frac{2 \gammae+\pi-2}{4 \pi} r \cos(\theta) 
% -\frac{r}{2 \pi} (\cos \theta +(2 \theta - \pi \Sign [\theta] )\sin \theta) -\frac{r^2}{2\sqrt{2 \pi}}  \Sign[\theta] \sin \theta + O(r^3),
\psi(r,\theta) = \frac{r \log r }{\pi} \cos \theta +\frac{2 \gammae+\pi-4}{4 \pi} r \cos \theta 
-\frac{r}{2 \pi} (2 \theta - \pi ~\Sign [\theta] )\sin \theta -\frac{r^2}{2\sqrt{2 \pi}}  \Sign[\theta] \sin \theta + O(r^3),
\end{equation}
where $\theta \in [-\pi,\pi]$ and $\gammae = 0.577\ldots$ is Euler's constant. Clearly 
the derivative $\partial_\theta \psi$ is discontinuous across $\theta=0$ and its value at $\theta=\pi$ is different from that at $\theta=-\pi$. 
Summing the three $\psi$ functions in \eqref{eq:C1psi}, one clearly sees that $\partial_\theta \cC_1$ 
is discontinuous along the three lines where the sources are located.
For larges values of $r$, the $\psi$ function concentrates around $\theta=0$ as it should do,
\begin{equation}
\label{eq:psilargerf}
\psi(r,\theta) = -\frac{1}{\sqrt{2 \pi}} \delta(\theta) \frac{\ee^{-\frac{r^2}{2}}}{r^4} \left(1+O\left(\frac{1}{r^2}\right)\right).
\end{equation}

% {\color{red}Asymptotics of $\cC_1(r,\theta)$ for  $r \to \infty$ and finite $\theta$ to be added}.

\begin{figure}[!ht]
	\begin{center}
		\includegraphics[width=0.5\textwidth]{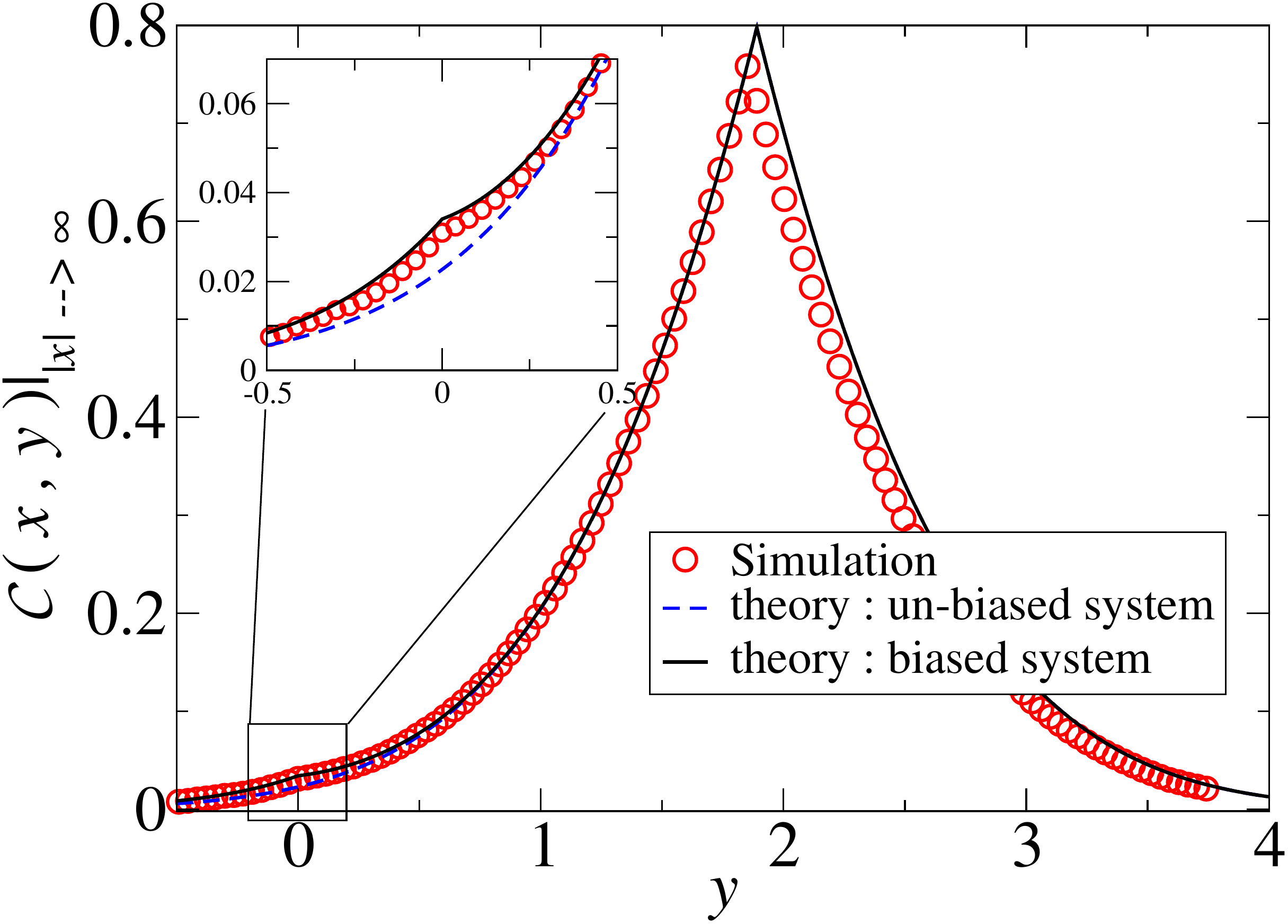} 
	\end{center}
	\caption{\small Plot of the theoretical $\cC(x,y)$ vs. $y$ for large $x$ (solid line) compared with numerical measurements (circles). Here $x=1.89$. The effect of 
                 the driven tracer is maximum near $y\sim 0$. To visualize this effect better, we zoomed in the region $y\sim 0$ in the inset. Other parameters 
                 associated with this plot are : $p=0.75$, $q=0.25$, $t=700$ and $N=200$. Jump distribution is uniform \emph{i.e.} $R(\eta)=1$.}  
	\label{fig:clrg}
\end{figure}

\subsubsection{Correlations for large $|x|$ }
\noindent
In the preceding section we have presented a perturbative method to find $\cC(x,y)$. This method is rather lengthy and cumbersome. However, when any one of the arguments 
(either $x$ or $y$) is large in magnitude, one can find explicitly a simpler approximate solution for $\cC(x,y)$. When, say $|x| \to \infty$, we simplify the diagonal source 
term on the RHS of \eqref{eq:evoCscalingsimp} by its large $x$ form  \emph{i.e.} put $\cH(x) \simeq 1$. As a result it now becomes easier to solve \eqref{eq:evoCscalingsimp} 
with boundary conditions \eqref{eq:evoCscalingbc} and \eqref{c-inf}. We find the following solutions when $|x| \to \infty$ :
\begin{eqnarray}
\label{C-x-inf}
\cC_{\infty}(x,y)= \cC(x,y)|_{|x| \to \infty} \simeq \frac{\mu_2}{\sqrt{2\pi}(\mu_1-\mu_2)}
\begin{cases}
 & g(x-y)+\frac{p-q}{p+q}~g(x+y),~~\text{for}~~x \geq 0,~~y\geq 0, \\
 & \frac{2p}{p+q}~g(x-y),~~~~~~~~~~~~~~~~~\text{for}~~x \geq 0,~~y < 0, \\
 & \frac{2q}{p+q}~g(x-y),~~~~~~~~~~~~~~~~~\text{for}~~x < 0,~~y \geq 0, \\
 &g(x-y)-\frac{p-q}{p+q}~g(x+y), ~~\text{for}~~x < 0,~~y < 0,
\end{cases}
\end{eqnarray}
where $g(u)$ is given in \eqref{g_u}. 
In fig.~\ref{fig:clrg} we compare these solutions with numerical results. The red circles represent data obtained from numerical simulation whereas 
black solid line represents the theoretical expression \eqref{C-x-inf}. For comparison we also have plotted the scaled 
correlation function $\cC_{ub}(x,y)$ corresponding to the unbiased system from \eqref{c-inf} (dashed blue line). 
We see that the effect of the biased tracer is maximum when $y$ is 
close to zero as expected. To visualize this effect better we zoomed in the region near $y \sim 0$ in the inset.
% While one particle is at large distance $x$, if one particle itself is close to the tracer 

\subsubsection{Special case : $q=0$ }
\label{q_0-case}
\noindent
This case is very interesting since for $q=0$ the particles in front of the biased tracer are not affected by the particles behind it.
As a result the boundary conditions in \eqref{eq:evoCscalingbc} becomes simpler :
\begin{eqnarray}
\label{bc-q_0}
 \partial_x \cC(0^+,y)=0,~~~\partial_y \cC(x,0^+)=0.
\end{eqnarray}
It turns out that now one can solve \eqref{eq:evoCscalingsimp} for $\cC(x,y)$ exactly in the first quadrant ($\cA_{++}=[x\ge 0,~y\ge 0]$) using image method. 
Before going into that let us look at the other boundary condition given in \eqref{c-inf}
which says that at distances far from the origin ( \emph{i.e.} far from the driven tracer ) the scaled correlation function should 
be the same as that of an unbiased system. 

At this point one would naturally intend to assume that $\cC(x,y)=\cC_{ub}(x,y)+ \bar{\cC}(x,y)$ and then solve for $\bar{\cC}(x,y)$. 
But this choice of decomposition of the solution 
is not useful since $\cC_{ub}(x,y)$ does not satisfy the boundary conditions \eqref{bc-q_0}. As a result it will make 
the boundary conditions for $\bar{\cC}(x,y)$ complicated. However, 
one can find a better decomposition 
\begin{equation}
 \cC(x,y)=\cC_{\infty}(x,y)+ \bar{\cC}(x,y) =  \frac{\mu_2}{\sqrt{2\pi}(\mu_1-\mu_2)}~[g(x-y)~+~g(x+y)] + \bar{\cC}(x,y) \label{decom-C-q-0}
\end{equation}
where $\cC_{\infty}(x,y)$ from \eqref{C-x-inf} with $q=0$ have been used. Note that $\cC_{\infty}(x,y)$ with $q=0$
satisfies both boundary conditions \eqref{bc-q_0} and \eqref{c-inf}. Hence the boundary conditions for $\bar{\cC}(x,y)$ remains the same, namely \eqref{bc-q_0} and \eqref{c-inf}. 
After inserting \eqref{decom-C-q-0} in \eqref{eq:evoCscalingsimp} 
%and using $\cH(x)=1-\text{Sign}(x)~\Erfc(|x|)$ from \eqref{eq:scalingH} for $q=0$, 
we have 
\begin{equation}
\label{Cbar_q-0-image}
[\partial_x^2 + \partial_y^2 +2 x \partial_x + 2 y \partial_y -2] \bar{\cC}(x,y) = -  \frac{2 \mu_2}{\mu_1-\mu_2} (\cH(x)^2-1) ~[\delta(x-y) + \delta(x+y)],
\end{equation}
% where the boundary conditions for $\bar{\cC}(x,y)$ remains the same, namely \eqref{bc-q_0} and \eqref{c-inf}. 

\begin{figure}[!ht]
	\begin{center}
		\includegraphics[width=0.5\textwidth]{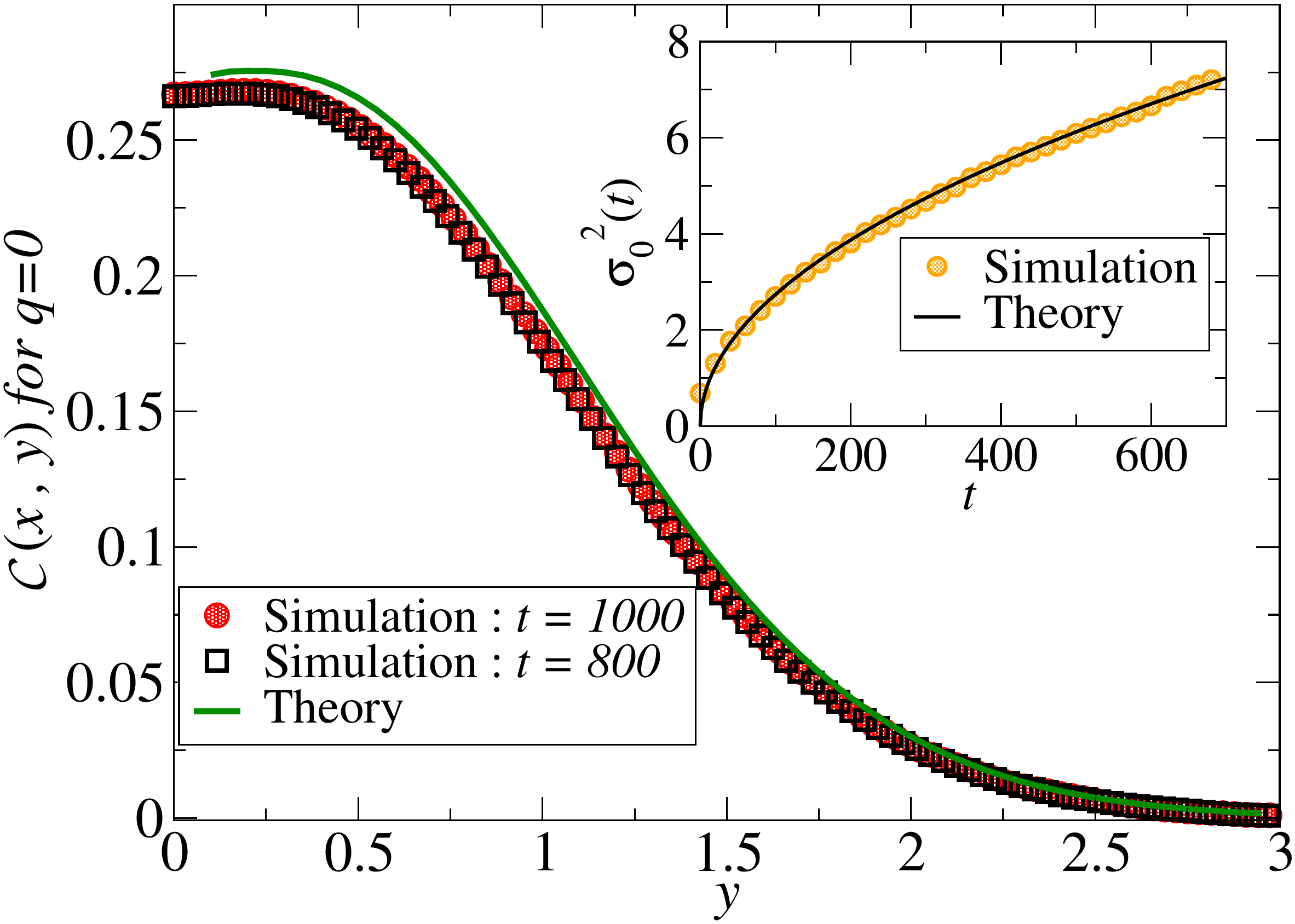} 
	\end{center}
	\caption{\small Theoretical $\cC(x,y)$ vs. $y$ for $q=0$ compared to numerical measurements. Here $x=0.01$. The theoretical curve (solid Green line) 
                is obtained using  equations \eqref{decom-C-q-0}, \eqref{eq:expxi} and \eqref{xi_m-r} where the infinite sum in \eqref{eq:expxi} has been truncated at 
                $m=4$. In the inset we compare the theoretical expression of $\sigma_0^2(t)$ in \eqref{sig_0_t} (solid Black line ) with numerical measurements 
                (Orange circles). Other parameters associated with this plot are : $L=1.0$ and $N=200$. Jump distribution is uniform \emph{i.e.} $R(\eta)=1$.}  
	\label{fig:c-q-0}
\end{figure}
We now proceed to solve \eqref{Cbar_q-0-image} for $(x,y)\in \cA_{++}$.
In this domain the sources (or 'charges') of the differential equation \eqref{Cbar_q-0-image} 
are distributed along $x=y$ line and the normal derivatives of $\bar{\cC}(x,y)$ at its boundaries ($x=0$ and $y=0$ line) vanish. 
Note that the 'charge' distribution along the diagonal in  $\cA_{++}$ is $\cH(x)^2-1=\Erf(x)^2-1$ [see \eqref{eq:scalingH} for $q=0$].
To solve the differential equation \eqref{Cbar_q-0-image} in $\cA_{++}$ with these boundary conditions we consider the following image problem : 
Since the differential operator $\hat{\cD}$ on the left hand side of \eqref{Cbar_q-0-image} is invariant under $x \to -x$ and/or $y\to -y$,
we consider the problem on complete two dimensional plane $\cA=[-\infty < x <\infty,~ -\infty < y <\infty]$ with three image 'charge' distributions obtained 
by reflecting the original 'charge' distribution with respect to the $x$-axis, $y$-axis and the origin respectively. As a result we automatically 
satisfy the boundary conditions in \eqref{bc-q_0} by symmetry. Hence we now solve 
\begin{equation}
\label{C_q-0-image}
[\partial_x^2 + \partial_y^2 +2 x \partial_x + 2 y \partial_y -2]\bar{\cC}(x,y) = -  \frac{2 \mu_2}{\mu_1-\mu_2} (\Erf(x)^2-1)~[\delta(x-y) + \delta(x+y)],
\end{equation}
in the full domain $\cA$, with the boundary conditions $\bar{\cC}(x,y)\to 0$ as $\sqrt{x^2+y^2} \to \infty$ and $\bar{\cC}(x,y)\to$ finite as $\sqrt{x^2+y^2} \to 0$. 
Once again going to the tilted polar coordinates $(x,y) = (r \cos \left( \theta + \frac{\pi}{4} \right), r \sin \left( \theta + \frac{\pi}{4} \right))$ 
(as done in \eqref{eq:evoC1scalingpol}) 
we rewrite \eqref{C_q-0-image} as 
\begingroup\makeatletter\def\f@size{9}\check@mathfonts
\def\maketag@@@#1{\hbox{\m@th\large\normalfont#1}}%
\begin{align}
\label{C_q-0-image-r}
 \left[\partial_r^2 +  \frac{1}{r} \partial_r + \frac{1}{r^2} \partial_\theta^2 +2 r \partial_r -2 \right] \bar{\cC}(r,\theta) 
= - \frac{\sqrt{2} \mu_2}{\mu_1-\mu_2} \frac{2}{\pi} \frac{(\Erf(r/\sqrt{2})^2-1)}{r} \left[\delta (\theta) + \delta \left(\theta-\frac{\pi}{2}\right)
+ \delta (\theta+\pi) + \delta \left(\theta+\frac{\pi}{2}\right)\right],~~~~
\end{align}
\endgroup
% \begin{eqnarray}
% \label{C_q-0-image-r}
%  \left[\partial_r^2 +  \frac{1}{r} \partial_r + \frac{1}{r^2} \partial_\theta^2 +2 r \partial_r -2 \right] \bar{\cC}(r,\theta) 
% = \small{- \frac{\sqrt{2} \mu_2}{\mu_1-\mu_2} \frac{2}{\pi} \frac{(\Erf(r/\sqrt{2})^2-1)}{r} \left[\delta (\theta) + \delta \left(\theta-\frac{\pi}{2}\right)
% + \delta (\theta+\pi) + \delta \left(\theta+\frac{\pi}{2}\right)\right]},~~~~
% \end{eqnarray}
where we have used explicit expression of $\cH(x)$ for $q=0$ from \eqref{eq:scalingH}.
To solve this equation we consider the following expansions : $\delta(\theta) = \frac{1}{2 \pi} [1+2\sum_{m=1}^\infty \cos(m \theta)]$ and 
\begin{equation}
\label{eq:expxi}
\bar{\cC}(r,\theta)  = - \frac{\sqrt{2} \mu_2}{\mu_1-\mu_2}\left(\xi_0(r)+\sum_{m=1}^\infty 2 \cos (4m\theta)~ \xi_m(r)\right).
\end{equation}
Inserting this form in \eqref{C_q-0-image-r} we find that the function $\xi_m(r)$ satisfies 
 \begin{eqnarray}
\label{eq:odexim}
&&\xi''_m(r) +  \left(\frac{1}{r} +2 r \right) \xi'_m(r) - \left(2+\frac{16m^2}{r^2}\right) \xi_m(r) = \frac{2}{\pi r}~[\Erf(r/\sqrt{2})^2-1], \label{src-q-0}
\end{eqnarray}
with boundary conditions 
\begin{equation}
\xi_0(r) |_{r \to 0} = \text{finite},~~\xi_{m>0}(r) |_{r \to 0} =0, \text{~~and~~~}  \xi_m(r) |_{r \to \infty} = 0~~~\forall m.
\end{equation}
Two homogeneous solutions of the above equation are :
\begin{eqnarray}
\label{xi-hom}
 \xi_m^{1h}(r)&=&r^{4 m} \, _1F_1\left(\frac{1}{2} (4 m-1);4 m+1;-r^2\right), \nonumber \\
 \xi_m^{2h}(r)&=&e^{-r^2} r^{-4 m} U\left(\frac{3}{2}-2 m,1-4 m,r^2\right),
\end{eqnarray}
where $_1F_1(a,b,z)$ is the hypergeometric function and $U(a,b,z)$ is Kummer hypergeometric function \cite{wolframfunctions}.
In terms of these homogeneous solutions the total solution is written as 
\begin{eqnarray}
% &&\xi_m(r)= \xi_m^{1h}(r) \int_r^\infty dr' \frac{\xi_m^{2h}(r')}{W(r',m)}~S(r') + \xi_m^{2h}(r) \int_0^r dr' \frac{\xi_m^{1h}(r')}{W(r',m)}~S(r'), \nonumber \\
&&\xi_m(r)= \xi_m^{1h}(r) \int_r^\infty dr' \frac{\xi_m^{2h}(r')}{W_m(r')}~\frac{2}{\pi r'}~[\Erf(r'/\sqrt{2})^2-1] 
+ \xi_m^{2h}(r) \int_0^r dr' \frac{\xi_m^{1h}(r')}{W_m(r')}~\frac{2}{\pi r'}~[\Erf(r'/\sqrt{2})^2-1], \label{xi_m-r} \\
&&~~~~~~~~~~~~~~~~~~~~~~~~~~~~~~~~~~~\text{where}~~W_m(r')= \xi_m^{1h}(r) \partial_r \xi_m^{2h}(r) - \xi_m^{2h}(r) \partial_r \xi_m^{1h}(r). \label{W_m}
\end{eqnarray}
One can numerically (in Mathematica) evaluate $\xi_m(r)$s as a function of $r$ for different values of $m$ and 
use them in \eqref{eq:expxi} to get $\bar{\cC}$ which finally provides $\cC$ in equation \eqref{decom-C-q-0}. In fig. \ref{fig:c-q-0} we compare the numerically evaluated 
$\cC$ using these equations 
% \eqref{decom-C-q-0}, \eqref{eq:expxi} and \eqref{xi_m-r}, 
with the simulation results. We observe good agreement with slight differences 
which arise because of the fact that the ring used in the simulation may be not in the thermodynamic limit.
% and $s(r)=\frac{2}{\pi r}~[\Erf(r/\sqrt{2})^2-1]$.

From the above analysis we can, in particular, compute the fluctuation of the displacement of the driven 
tracer particle $\sigma_0^2(t)=\langle x_0^2(t) \rangle -\langle x_0(t) \rangle^2$ in large time limit. 
In terms of the scaled correlation function this quantity is given by $\sigma_0^2(t)=\rho_0^{-2}\sqrt{2\mu_1 t}~\cC(0,0)$. More explicitly we have  
\begin{equation}
 \cC(0,0)=\frac{\mu_2}{\sqrt{2\pi}(\mu_1-\mu_2)}-\frac{\sqrt{2} \mu_2}{\mu_1-\mu_2}~I,
 ~~\text{where}~I=\int_0^\infty dr' \frac{\xi_0^{2h}(r')}{W_0(r')}~\frac{2}{\pi r'}~[\Erf(r'/\sqrt{2})^2-1], \label{I}
\end{equation}
with $\xi_0^{2h}(r)$ and $W_0(r)$ given in \eqref{xi-hom} and \eqref{W_m} respectively. One can perform this integral exactly (see appendix \ref{app-I} )to get  
$I=\frac{2}{\sqrt{\pi}}(\sqrt{2}-1)$ which implies
\begin{equation}
 \sigma_0^2(t)= \rho_0^{-2} \frac{\mu_2}{(\mu_1-\mu_2)}~\sqrt{\frac{2}{\pi}}~\frac{\sqrt{2}-1}{\sqrt{2}+1}~\sqrt{2\mu_1 t}. \label{sig_0_t}
\end{equation}
In the inset of fig. \ref{fig:c-q-0} we verify this result numerically. 

% \noindent
In the context of SSEP with driven tracer, similar late time growth $\sim \sqrt{t}$ for 
the fluctuation of the tracer position have been reported~\cite{Illien-13, Burlatsky-92}. Moreover, other moments of the tracer position 
have also been computed using different approximations. For example, in~\cite{Illien-13} SSEP with driven tracer have been studied in the high density regime. 
In particular, the authors of~\cite{Illien-13} have found the distribution of the position of the biased particle by mapping the motion of the particles to appropriate 
random walks of the holes (absence of particles) for the high particle density. However, none of the studies have considered computing the position pair correlation function. 
To our knowledge, the calculation in this paper is the first attempt for computing such correlation function without any approximation.

\subsection{Gap correlations}
\label{gap-corr}
\noindent
Let us now focus on the pair gap correlation $d_{i,j}(t)$ defined in \eqref{eq:defdij}. To compute the evolution equation for $d_{i,j}(t)$, it is convenient 
to consider the dynamics of the stochastic gap variables $g_i(i)=x_{i+1}(t)-x_i(t)$ independently. From the dynamics \eqref{dyna-x} of $x_i(t)$s, one can write the dynamics 
of $g_i(t)$s as 
\begin{equation}
g_i(t+dt) = g_i(t) + \sigma^{i+1}_r \eta_{i+1} g_{i+1}(t) + \sigma^{i}_l \eta_{i} g_{i-1}(t) 
- \sigma^{i+1}_l \eta_{i+1} g_{i}(t) - \sigma^{i}_r \eta_{i} g_{i}(t), \label{dyna-gap}
\end{equation} 
where the random fraction $\eta$ is chosen from jump distribution $R(\eta)$ and the variables $\sigma_{l,r}^i$ are defined after \eqref{dyna-x}.
Using this dynamics, it is straightforward to find evolution equations for $d_{i,j}$,
\begin{eqnarray}
\label{eq:evodij}
\dot{d}_{i,j} &=& \frac{\mu_1}{2} (d_{i+1,j} + d_{i-1,j} + d_{i,j+1} + d_{i,j-1} - 4 d_{i,j}) \nonumber \\
&&+ \frac{\mu_2}{2} (\delta_{j,i} - \delta_{j,i+1}) (d_{i+1,i+1} + d_{i,i} + h_{i+1}^2 + h_i^2) 
+ \frac{\mu_2}{2} (\delta_{j,i} - \delta_{j,i-1}) (d_{i-1,i-1} + d_{i,i} + h_{i-1}^2 + h_i^2) \\
&&+ (\delta_{i,-1}-\delta_{i,0}) \frac{\mu_1}{2} ((2p-1) d_{j,0} - (2q-1) d_{j,-1}) 
+ (\delta_{j,-1}-\delta_{j,0}) \frac{\mu_1}{2} ((2p-1) d_{i,0} - (2q-1) d_{i,-1}) \nonumber \\
&&+ (\delta_{i,-1}-\delta_{i,0}) (\delta_{j,-1}-\delta_{j,0}) \frac{\mu_2}{2} ((2q-1) (d_{-1,-1}+h_{-1}^2)+(2p-1) (d_{0,0}+h_{0}^2)), \nonumber
\end{eqnarray}
where $h_i(t)=\langle g_i(t)\rangle$. 
\begin{figure}[!ht]
	\begin{center}
		\includegraphics[width=0.5\textwidth]{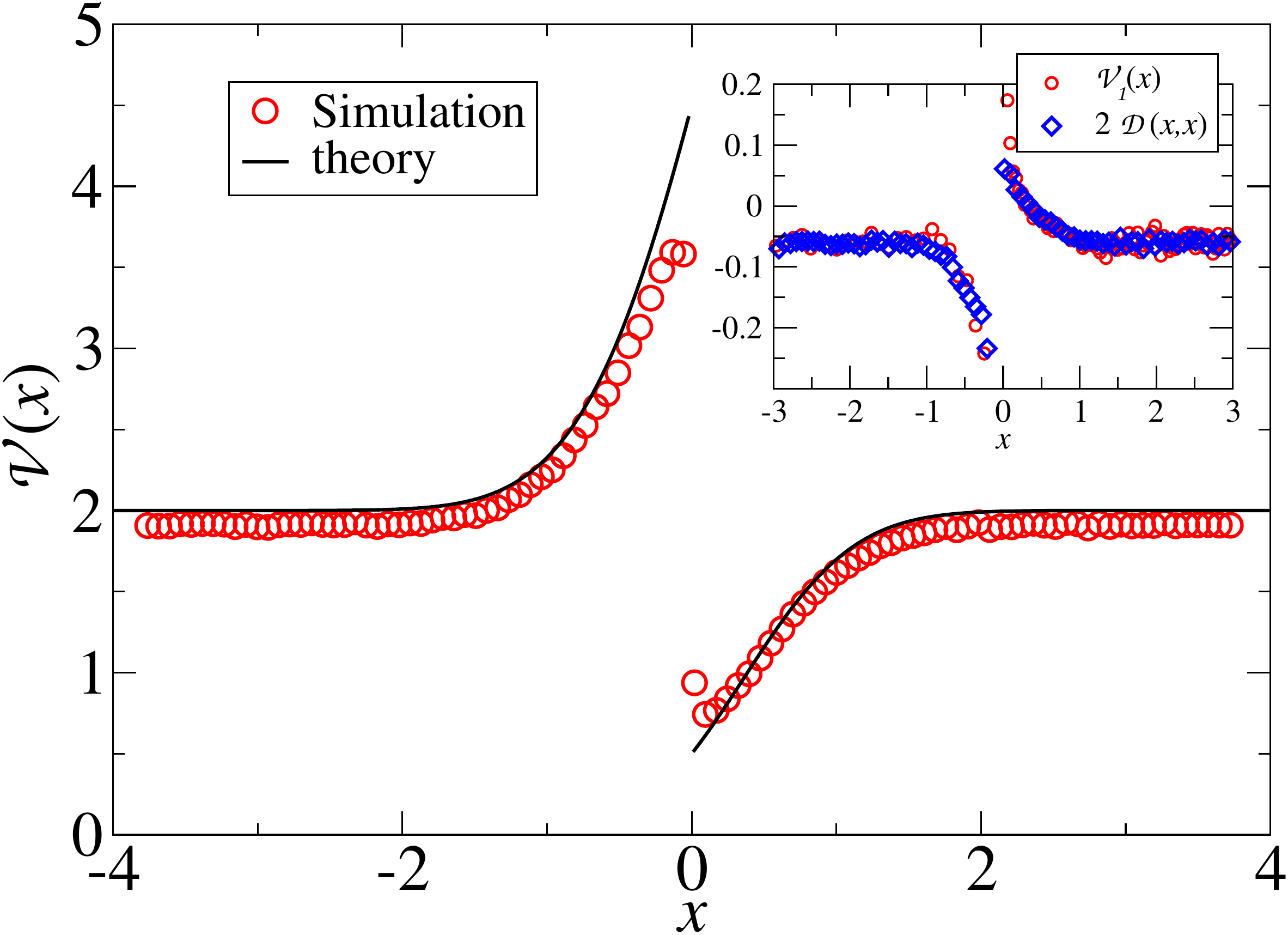} 
	\end{center}
	\caption{\small Numerical verification of \eqref{eq:solV} and \eqref{eq:solV1} (inset).
                  Numerical values for $\cD(x,x)$ in simulation are obtained from $d_{j-1,j+1}(t)$. 
                 The parameters associated with this plot are : $p=0.75$, $q=0.25$, $t=700$ and $N=200$. 
                 Jump distribution is uniform \emph{i.e.} $R(\eta)=1$.}  
	\label{fig:V}
\end{figure}
In the beginning of section \ref{section:2pt}, 
we argued that in large time limit diagonal and off-diagonal gap correlations scale differently as \eqref{eq:scalingdd} and \eqref{eq:scalingd}
respectively. We are interested in finding solutions of $d_{i,j}(t)$ in these scaling forms. Once again note that, while off-diagonal correlations 
are of order $t^{-1/2}$ the diagonal correlations \emph{i.e.} fluctuations are order one. We now insert the scaling forms of $d_{i,j}(t)$ from 
\eqref{eq:scalingd}, $d_{i,i}(t)$ from \eqref{eq:scalingdd} and $h_i(t)$ from \eqref{eq:gapscaling} in \eqref{eq:evodij} and then expand both sides in powers of 
$\frac{1}{\sqrt{t}}$. Equating coefficients of each powers from both sides, we find that orders $t^{-1/2}$ and $t^{-1}$ give
\begin{eqnarray}
\cV(x) &=& \frac{\mu_2}{\mu_1 - \mu_2} \cH(x)^2 = \frac{\mu_2}{\mu_1 - \mu_2} \left[ 1 - \frac{p-q}{p+q} \Sign(x) \Erfc\left(|x|\right) \right]^2, \label{eq:solV} \\
&&~~~~~~\cV_1(x) = \frac{\mu_2}{\mu_1 - \mu_2} \cD(x,x), \label{eq:solV1}
\end{eqnarray}
where $\cD(x,y)$ is completely determined from the knowledge of $\cC(x,y)$ through \eqref{eq:linkcd}. In fig.~\ref{fig:V} we numerically verify \eqref{eq:solV} 
whereas in the inset we verify \eqref{eq:solV1}. For the plot in the inset, both the quantities$\cV_1(x)$ and $\cD(x,x)$ are obtained from numerical measurements. 
Numerical values for $\cD(x,x)$ in simulation are obtained from $d_{j-1,j+1}(t)$.

\section{Conclusion}
\label{conclusion}
\noindent
In this work we studied the motion of a driven tracer particle in an otherwise symmetric Random Average Process on an infinite line. 
In the first part, the motion of the tracer and its effects on its environment have been characterized by computing the displacement of the 
tracer as well as the perturbation of the density profile. For both quantities the results are very similar to those obtained in a 
one dimensional SSEP with a single biased tracer, where the velocity of the tracer also vanishes at large times and 
the density perturbation decays exactly the same way at large distances.

In single file systems particles are subjected to strong caging effects, 
which usually have dramatic effects on the fluctuations and correlations of the positions of the particles. 
Since in our case the particles are also non-overtaking, their motion constitutes a single file motion. We have shown in this paper that, 
at large times, the position-position correlations of different particles at equal time support nice scaling form when the particle labels are rescaled by $\sqrt{t}$. 
We showed that the corresponding scaling function $\cC(x,y)$ satisfies some differential equation which can be solved perturbatively around the solution of the unbiased tracer case. 
We have computed the first two terms of the perturbative expansion. In the case where the tracer is totally asymmetric the problem is more tractable, enabling us to compute 
the variance of the position of the tracer exactly. Finally, the variances of the gaps between successive particles were obtained and 
shown to converge to finite values at large times.

% Only a few results concerning the fluctuations and correlations of the positions of the particles have been obtained in this paper. 
There are many interesting 
extensions of this problem to explore in future. For example, finding an exact and complete solution of the equation \eqref{eq:evoCscalingsimp} for arbitrary 
$p$ and $q$ would be of interest. In equations \eqref{pos-indx-rela} and \eqref{den-scling} we have obtained the average density profile. However, the fluctuations of the local density about this average remain to be calculated.
%But, what are the fluctuations 
%of the local density about this average value?
%In fact, finding the probability of observing a deviation of the full density profile from the average profile is of high interest 
%currently. Also people have recently been interested in the large deviation function associated with the distribution of the position of the driven tracer particle at 
%arbitrary but large time. Studying such large deviation function would also be an important future extension of our work.
Also, calculating the probability of large deviations, either of the full density profile or the position of the driven tracer, would be of interest.
A different problem which one would like to explore is the effective interaction between two or more tracers and the dynamical effect that one tracer has on another.

\section{Acknowledgements}
\noindent
We thank Victor Mukherjee for carefully reading the manuscript. The support of the Israel Science Foundation
(ISF) and of the Minerva Foundation with funding from the Federal German Ministry for Education and Research
is gratefully acknowledged. S.N.M wants to thank the hospitality of the Weizmann Institute where this work started
during his visit as a Weston visiting professor.
% {\color{red}{This section needs to be improved more.}}
%
\appendix

\section{Solution of $\psi(r,\theta)$}
\label{section:appC1}

% \subsection{Expansion of the angular part}

In this appendix we solve the equation~\eqref{eq:psi} for the function $\psi(r,\theta)$ in polar coordinates. For convenience let us rewrite \eqref{eq:psi} here :
\begin{equation}
\label{eq:psi1}
\left[\partial_r^2 +  \frac{1}{r} \partial_r + \frac{1}{r^2} \partial_\theta^2 +2 r \partial_r -2 \right] \psi(r,\theta) 
=  \left(\delta\left(\theta \right)-\delta \left( \theta-\pi \right)\right) \frac{\Erfc(r/\sqrt{2})}{r}.
\end{equation}
As we noticed in \eqref{eq:C1psi} that $\cC(r, \theta)$ can be expressed in terms of $\psi(r,\theta)$ as 
\begin{equation}
\label{eq:C1psi0}
\cC_1(r,\theta) = \frac{2 \mu_2}{\mu_1-\mu_2} \left( 2 \sqrt{2} \psi(r,\theta) - \psi\left(r,\theta + \frac{\pi}{4}\right) 
- \psi\left(r,\theta - \frac{\pi}{4}\right) \right).
\end{equation}
We now use the representation of the delta function $\delta(\theta) = \frac{1}{2 \pi} \sum_{l=-\infty}^\infty \ee^{\ci l \theta}$ to write
\begin{equation}
\label{eq:expdelta}
\delta\left(\theta \right)-\delta \left( \theta-\pi \right) = \frac{1}{\pi} \sum_{m=0}^\infty 2 \cos ((2m+1)\theta),
\end{equation}
and to expand the angular part of $\psi$ as well, 
\begin{equation}
\label{eq:exppsi1}
\psi(r,\theta) = \sum_{m=0}^\infty 2 \cos ((2m+1)\theta) \psi_m(r).
\end{equation}
Using this expansion on both sides of \eqref{eq:psi1}, we find 
\begin{equation}
\label{eq:odepsip}
\psi_m''(r) +  \left(\frac{1}{r} +2 r \right) \psi_m'(r) - \left(2+\frac{(2m+1)^2}{r^2}\right) \psi_m(r) = \frac{\Erfc(r/\sqrt{2})}{\pi r},
\end{equation}
with the conditions that $\psi_m(r)$ is finite for $r \rightarrow 0$ and vanishes for $r \rightarrow \infty$.

\subsection{Solution for $m\geq1$}
\noindent
For $m \geq 1$, the general solution of \eqref{eq:odepsip} reads
\begin{equation}
\label{eq:solpsip}\psi_m(r) = C^1_m \psi^1_m(r) + C^2_m \psi^2_m(r) + \psi^P_m(r),
\end{equation}
where $C^1_m$ and $C^2_m$ are constants to be determined, $\psi^1_m(r)$ and $\psi^2_m(r)$ are solutions of the homogeneous equation,
\begin{eqnarray}
\label{eq:psip12}
\psi^1_m(r) &=& \frac{\ee^{-r^2}}{r^{2 m+1}} \sum_{k=0}^{m-1} \left( \prod_{l=k}^{m-2} \frac{(l+1) (l-2 m)}{l+1-m}\right) r^{2 k}, \\
\psi^2_m(r) &=& \frac{1}{r^{2 m+1}} \sum_{k=0}^{m+1} \left( \prod_{l=k}^{m} \frac{(l+1) (l-2 m)}{m+1-l}\right) r^{2 k}, \nonumber
\end{eqnarray}
and $\psi^P_m(r)$ is a particular solution. Looking at the structure of the soultions for small values of $m$, we try a particular solution of the form
\begin{eqnarray}
\label{eq:formpsip}
\psi^P_m(r) &=& \frac{\ee^{-r^2}}{r^{2 m+1}} \chi_m(r), \qquad~\mathrm{with} \\
\chi_m(r) &=& \ee^{\frac{r^2}{2}} P_{1,m}(r) + \ee^{r^2} \Erf\left( \frac{r}{\sqrt{2}} \right) P_{2,m}(r) 
+ \ee^{r^2} \Erfc\left( \frac{r}{\sqrt{2}} \right) P_{3,m}(r) + \Erfi\left( \frac{r}{\sqrt{2}} \right) P_{4,m}(r),\nonumber
\end{eqnarray}
where $\Erf(z) = \frac{2}{\sqrt{\pi}} \int_{t=0}^z \ee^{-t^2} \dd t$ is the error function, $\Erfi(z) = \frac{2}{\sqrt{\pi}} \int_{t=0}^z \ee^{t^2} \dd t$ 
is the 'imaginary error function' and the $P_{i,m}(r)$s are polynomials that depend on $m$, although this is not emphasized by the notation. 
Substituting $\chi^P_m(r)$ in \eqref{eq:odepsip} we have
\begin{equation}
\label{eq:odechip}
\chi_m''(r) - \left( \frac{1+4m}{r} +2r\right) \chi_m'(r) + 4 (m-1) \chi_m(r) = \frac{r^{2m}}{\pi} \ee^{r^2} \Erfc\left( \frac{r}{\sqrt{2}} \right).
\end{equation}
The exponential functions in the prefactors of the polynomials in equation~\eqref{eq:formpsip} cannot be generated by polynomials. Equation~\eqref{eq:odechip} 
is therefore verified iff the coefficient of each of the functions 
$\ee^{\frac{r^2}{2}}$,$\ee^{r^2} \Erf\left( \frac{r}{\sqrt{2}} \right)$,$\ee^{r^2} \Erfc\left( \frac{r}{\sqrt{2}} \right)$ and $\Erfi\left( \frac{r}{\sqrt{2}} \right)$ 
vanishes. The $\ee^{r^2} \Erf\left( \frac{r}{\sqrt{2}} \right)$,$\ee^{r^2} \Erfc\left( \frac{r}{\sqrt{2}} \right)$ and $\Erfi\left( \frac{r}{\sqrt{2}} \right)$ 
involve only $P_{2,m}$, $P_{3,m}$ and $P_{4,m}$ respectively. The equations are
\begin{eqnarray}
\label{eq:odePi234}
P_{2,m}'' + \left(2 r - \frac{1+4m}{r}\right)P_{2,m}' -4 \left( p+1\right) P_{2,m} &=& 0, \nonumber \\
P_{3,m}'' + \left(2 r - \frac{1+4m}{r}\right)P_{3,m}' -4 \left( p+1\right) P_{3,m} &=& \frac{r^{2p}}{\pi}, \\
P_{4,m}'' - \left(2 r + \frac{1+4m}{r}\right)P_{4,m}' -4 \left( p-1\right) P_{4,m} &=& 0. \nonumber
\end{eqnarray}
Polynomial solutions of equations~\eqref{eq:odePi234} are easily found,
\begin{eqnarray}
\label{eq:solPi234}
P_{2,m}(r) &=& \frac{K_{2,m}}{4 \pi} \sum_{k=0}^{m+1} \left( \prod_{l=k+1}^{m+1} \frac{l (l-2m-1)}{m+2-l}\right) r^{2k}, \nonumber \\
P_{3,m}(r) &=& -\frac{1}{4 \pi} \sum_{k=0}^{m} \left( \prod_{l=k+1}^{m} \frac{l (l-2m-1)}{m+2-l}\right) r^{2k}, \\
P_{4,m}(r) &=& \frac{K_{4,m}}{4 \pi} \sum_{k=0}^{m-1} \left( \prod_{l=k+1}^{m-1} \frac{l (l-2m-1)}{l-m}\right) r^{2k}, \nonumber
\end{eqnarray}
where $K_{2,m}$ and $K_{4,m}$ are \textit{a priori} arbitrary constants that depend on $m$.

\noindent
Now we focus on the equation for $P_{1,m}$ which involves the other three polynomials,
\begin{eqnarray}
\label{eq:odePi1}
P_{1,m}'' - \frac{1+4m}{r} P_{1,m}' - (4+r^2) P_{1,m} &=&-\sqrt{\frac{2}{\pi}} \Big{[} 2 P_{4,m}' 
- \left( \frac{1+4m}{r}+r \right) P_{4,m}  \nonumber \\
&&~~~~+ 2 (P_{2,m}'-P_{3,m}') - \left( \frac{1+4m}{r}-r \right) (P_{2,m}-P_{3,m})\Big{]}.
\end{eqnarray}
We take $P_{1,m}(r) = \sum_{k=0}^{m} f_{2k+1} r^{2k+1}$. Identifying the powers of $r$ in~\eqref{eq:odePi1} gives $m+3$ equations 
for the $m+1$ coefficients $f_{2k+1}$, to which we add the unknown constants $K_{2,m}$ and $K_{4,m}$. The equations are clearly linear in 
the $f_{2k+1}$ and in $K_{2,m}$, $K_{4,m}$, so that they may be solved by matrix inversion. 

\noindent
There is no simple expression of the inverse matrix, but based on numerical solutions (found using Mathematica) for first few values of $m$, it seems reasonable to assume that 
equation~\eqref{eq:odePi1} has a unique solution in terms of the $f_{2k+1}$, $K_{2,m}$ and $K_{4,m}$. 
The polynomials $P_{1,m}$ for the first few values of $m$ are given as : 
\begin{eqnarray}
\label{eq:solP1}
P_{1,1}(r) &=& \frac{\sqrt{2} r}{4 \pi^{3/2}} \left( r^2 -2 \right), \nonumber \\
P_{1,2}(r) &=& \frac{\sqrt{2} r^3}{12 \pi^{3/2}} \left( r^2 - 4 \right), \nonumber \\
P_{1,3}(r) &=& \frac{\sqrt{2} r}{4 \pi^{3/2}} \left( r^4-10 r^2 +60 \right), \nonumber \\
P_{1,4}(r) &=& \frac{\sqrt{2} r^3}{4 \pi^{3/2}} \left( r^4 -14 r^2 +112 \right). \nonumber \\
P_{1,5}(r) &=& \frac{\sqrt{2} r}{60 \pi^{3/2}} \left( r^{10}-16 r^8+168 r^6-2520 r^4+15120 r^2-151200\right), \\
P_{1,6}(r) &=& \frac{\sqrt{2} r}{84 \pi^{3/2}} \left( r^{10}-22 r^8+360 r^6-6840 r^4+55440 r^2-665280 \right), \nonumber \\
P_{1,7}(r) &=& \frac{\sqrt{2} r}{4 \pi^{3/2}} \left( r^{12}-32 r^{10}+610 r^8-4752 r^6+78408 r^4-308880 r^2+4324320 \right), \nonumber \\
P_{1,8}(r) &=& \frac{\sqrt{2} r^3}{4 \pi^{3/2}} \left( r^{12}-40 r^{10}+970 r^8-11440 r^6+223080 r^4-1201200 r^2+19219200 \right). \nonumber
%\\
%&\vdots& \nonumber
\end{eqnarray} 
We sued these explicit forms of the polynomials to generate the theoretical curves 
in figures~\ref{fig:Csc} and~\ref{fig:C1}.
Based on small values of $m$, the values of the constants are conjectured to be
\begin{eqnarray}
\label{eq:solK2K4}
K_{2,m} &=&
\begin{cases}
& \frac{2}{m (m+1)} \qquad \mathrm{for}~ m=4l+1~\mathrm{or}~p=4l+2 \\
& 0 \qquad ~~~~~~~~\mathrm{for}~ m=4l+3~\mathrm{or}~m=4l+4
\end{cases}, \nonumber \\
&&~~~~~~~~~~~~~~~~~~~~~~~~~~~~~~~~~~~~~~~~~~~~~~~~~~~~~~~~~~~~~~~~~~~~~~~\text{for}~~l=0,1,2,\ldots \\
K_{4,m} &=&
\begin{cases}
& 1 \qquad ~~~~~~~~\mathrm{for}~ m=4l+1~\mathrm{or}~m=4l+2 \\
& -1 \qquad ~~~~~~\mathrm{for}~ m=4l+3~\mathrm{or}~m=4l+4
\end{cases}, \nonumber
\end{eqnarray}
which have been checked up to $m=8$.
Inserting the expressions of the polynomials from \eqref{eq:solPi234}-\eqref{eq:solP1} and the constants from \eqref{eq:solK2K4}, in the Ansatz \eqref{eq:formpsip} one obtains the 
particular solution $\psi_m^P(r)$.
% Combining the Ansatz~\eqref{eq:formpsip} for $\psi_m^P(r)$, the expressions of the polynomials~\eqref{eq:solPi234}-\eqref{eq:solP1} 
% and those of the constants~\eqref{eq:solK2K4} one obtains the particular solution.

\noindent
Let us now fix the constants $C_m^1$ and $C_m^2$. 
For large $r$ the particular solution goes like $\psi_m^P(r) \sim \frac{K_{2,m}}{4 \pi} r$, which must be compensated by $C_m^2 r$, giving 
\begin{equation}
 C_m^2 = - \frac{K_{2,m}}{4 \pi}.
\end{equation}
For small $r$ the homogeneous terms go like $\left(C_m^1 \frac{(2m)!}{(m+1)!} + C_m^2 \frac{(-1)^{m+1} (2m)!}{(m-1)!} \right) r^{-2m-1}$ 
and the particular solution like $\frac{1}{4 \pi} \frac{(-1)^{m+1} (2m)!}{(m+1)!}$, giving 
\begin{equation}
 C_m^1 = \frac{(-1)^{m+1}}{4 \pi}( m (m+1) K_{2,m}-1).
\end{equation}

%[The equation for $P_1$ seems OK. Numerically:
%\begin{itemize}
%	\item $K_2$ seems to be $\frac{2}{p (p+1)}$ for $p=4k+1$ and $p=4k+2$, $0$ for $p=4k+3$ and $p=4k+4$
%	\item $K_4$ seems to be $1$ for $p=4k+1$ and $p=4k+2$, $-1$ for $p=4k+3$ and $p=4k+4$
%	\item The degree of $P_1$ is $2p+1$ for $p=4k+1$ and $p=4k+2$, $2p-1$ for $p=4k+3$ and $p=4k+4$
%	\item The lowest order term of $P_1$ is $r$ for odd $p$ and $r^3$ for even $p$
%\end{itemize}
%
%]

\subsection{Solution for $m=0$}

For $m=0$ the general $m$ calculation does not hold, as the sum in the definition of $\psi_m^1$ would be empty. One can however find the homogeneous solutions separately as 
\begin{eqnarray}
\label{eq:psi012}
\psi^1_0(r) &=& \frac{\ee^{-r^2} + r^2 \Ei(-r^2)}{r}, \\
\psi^2_0(r) &=& r, \nonumber
\end{eqnarray}
where $\Ei(u) = - \int_{t=-u}^\infty \frac{\ee^{-t}}{t} \dd t$ is the exponential integral. The particular solution can be obtained from these homogeneous solutions as
\begin{equation}
\label{eq:psi0p}
\psi^P_0(r) = \int_{r'=1}^r \frac{\psi^1_0(r') \psi^2_0(r)-\psi^1_0(r) \psi^2_0(r')}{{\psi^1_0}'(r') \psi^2_0(r')
-\psi^1_0(r') {\psi^2_0}'(r')} \frac{1}{\pi r'} \Erfc\left( \frac{r'}{\sqrt{2}}\right) \dd r'.
\end{equation}

When $r$ is small we have $\psi_0^P(r) \sim -\frac{1}{4 \pi r}$, which must be compensated by $C^1_0 \psi_0^1(r) \sim \frac{C^1_0}{r}$, 
giving $C^1_0 = \frac{1}{4 \pi}$. When $r \rightarrow \infty$ we have
\begin{equation}
\label{eq:psP0inf}
 \psi_0^P(r) \sim r \int_{u=1}^\infty \left( 1+ u^2 \ee^{u^2} \Ei(-u^2)\right) \frac{1}{2 \pi u} \Erfc \left( \frac{u}{\sqrt{2}}\right) \dd u,
\end{equation}
that has to be compensated by $C^2_0 \psi^2_0(r) = C^2_0 r$, giving $C^2_0 = 
- \int_{1}^\infty \left( 1+ u^2 \ee^{u^2} \Ei(-u^2)\right) \frac{1}{2 \pi u} \Erfc \left( \frac{u}{\sqrt{2}}\right) \dd u \simeq -0.00589612 $.

The function $\psi(r,\theta)$ is obtained after summing~\eqref{eq:exppsi} over $m$ using $\psi_m(r,\theta)$ from \eqref{eq:solpsip} and using this $\psi(r,\theta)$ 
the solution $\cC_1(r,\theta)$ is obtained from \eqref{eq:C1psi}.

\subsection{Small and large $r$ behavior of $\psi(r,\theta)$}
\label{sml_lrg_r-psi}
\noindent
As a matter of fact the above choices of constants $C^1_m$ and $C^2_m$ ensure that, not only the strongest divergence of $\psi_m(r)$ gets canceled, 
but that the $\psi_m(r)$ vanish when $r \rightarrow 0$. Here we determine an expansion of $\psi(r,\theta)$ around $r=0$.
Each of the $\psi_m(r)$ may be expanded separately. For $m=0$ we get
\begin{equation}
\label{eq:psi0smallr}
\psi_0(r) = \frac{r \log r}{2 \pi} +\frac{2 \gammae + \pi -2}{8 \pi} r -\frac{\sqrt{2}}{3 \pi^3/2} r^2 +O(r^3),
\end{equation}
where $\gammae$ is Euler's constant. On numerical basis we can conjecture a general form of the expansion for any $m \geq 1$,
\begin{equation}
\label{eq:psimsmallr}
\psi_m(r) = - \frac{r}{4 \pi m(m+1)} + \frac{\sqrt{2}}{\pi^{3/2}} \frac{r^2}{4 (m+1)m-3}+O(r^3).
\end{equation}
The small $r$ behavior of $\psi(r,\theta)$ is therefore given by
\begin{equation}
\label{eq:psismallr}
\psi(r,\theta) = \frac{r \log r }{\pi}\cos \theta +\frac{2 \gammae+\pi-2}{4 \pi} r \cos \theta -\frac{r}{2 \pi} S_1(\theta) +\left( \frac{2}{\pi} \right)^{3/2} r^2 S_2(\theta) + O(r^3),
\end{equation}
where the sums are given by
\begin{eqnarray}
\label{eq:sumstheta}
S_1(\theta) &=& \sum_{m=1}^\infty \frac{\cos((2m+1)\theta)}{m (m+1)} = \cos \theta +(2 \theta - \pi \Sign [\theta] )\sin \theta \nonumber \\
S_2(\theta) &=& \sum_{m=0}^\infty \frac{\cos((2m+1)\theta)}{4 m (m+1)-3} = -\frac{\pi}{8} \Sign[\theta] \sin \theta,
\end{eqnarray}
for $-\pi < \theta \leq \pi$. Combining~\eqref{eq:psismallr} and \eqref{eq:sumstheta} we get equation~\eqref{eq:psismallrf} as presented in the main text.

For $r \rightarrow \infty$ it can be shown that all the $\psi_m$ functions behave like $-\frac{\sqrt{2}}{\pi^{3/2}} \frac{\ee^{-\frac{r^2}{2}}}{r^4}$. 
The radial part of the $\psi_m$ functions can be factorized out of the sum over $m$ and the angular part gives back a Dirac delta, 
giving expression~\eqref{eq:psilargerf} from the main text.

\section{Evaluation of the integral $I$ in \eqref{I}}
\label{app-I}
\noindent
Here we perform the integral
\begin{equation}
I=\int_0^\infty dr \frac{\xi_0^{2h}(r)}{W_0(r)}~\frac{2}{\pi r}~[\Erf(r/\sqrt{2})^2-1], \label{I-app-1}
\end{equation}
exactly. For $m=0$ the homogeneous solutions are more explicitly written as 
\begin{eqnarray}
\label{xi-hom-m-0}
 \xi_0^{1h}(r)&=&e^{-\frac{r^2}{2}} \left[r^2 I_1\left(\frac{r^2}{2}\right)+\left(r^2+1\right) I_0\left(\frac{r^2}{2}\right)\right], \nonumber \\
 \xi_0^{2h}(r)&=& \frac{2 }{\sqrt{\pi }}e^{-\frac{r^2}{2}} \left[\left(r^2+1\right) K_0\left(\frac{r^2}{2}\right)-r^2
   K_1\left(\frac{r^2}{2}\right)\right],
\end{eqnarray}
where $I_n(x)$ and $K_n(x)$ are modified Bessel functions of, respectively, the first and the second kind of order $n$.
Using the expressions of the derivatives of the Bessel functions in terms of Bessel functions of higher order and the recurrence 
relation between successive Bessel functions, the Wronskian can be brought to a very simple form, 
\begin{eqnarray}
\label{eq:W0simp}
W_0(r) &=& \frac{4}{\sqrt{\pi}} \frac{\ee^{-r^2}}{r}.
\end{eqnarray}
Hence simplifying \eqref{I-app-2} we have :
\begin{equation}
I=\frac{1}{\pi}\int_0^\infty dr e^{\frac{r^2}{2}} \left[\left(r^2+1\right) K_0\left(\frac{r^2}{2}\right)-r^2
   K_1\left(\frac{r^2}{2}\right)\right] ~[\Erf(r/\sqrt{2})^2-1], \label{I-app-2}
\end{equation}
Next, we take benefit of the following identity
\begin{equation}
\label{eq:dK0exp}
\frac{\dd}{\dd r} \left( \ee^{\frac{r^2}{2}} K_0\left( \frac{r^2}{2}\right)\right) 
= r \ee^{\frac{r^2}{2}} \left( K_0\left(\frac{r^2}{2}\right)-K_1\left(\frac{r^2}{2}\right)\right)
\end{equation}
to integrate by parts
\begin{eqnarray}
 \label{I-app-3}
I&=&\frac{1}{\pi}\int_{r=0}^\infty \dd r e^{\frac{r^2}{2}} \left(\left(r^2+1\right) K_0\left(\frac{r^2}{2}\right)-r^2
   K_1\left(\frac{r^2}{2}\right)\right) ~\left(\Erf\left(\frac{r}{\sqrt{2}}\right)^2-1\right) \nonumber \\
   &=& \frac{1}{\pi} \int_{r=0}^\infty \dd r \ee^{\frac{r^2}{2}} K_0\left( \frac{r^2}{2}\right) \left[\Erf\left(\frac{r}{\sqrt{2}}\right)^2-1\right] 
+ \frac{1}{\pi} \left[ r \ee^{\frac{r^2}{2}} K_0\left( \frac{r^2}{2}\right) \left(\Erf\left(\frac{r}{\sqrt{2}}\right)^2-1\right)\right]\Bigg|_{r=0}^\infty \nonumber \\
&& - \frac{1}{\pi} \int_{r=0}^\infty \ee^{\frac{r^2}{2}} K_0\left( \frac{r^2}{2}\right) 
\frac{\dd}{\dd r} \left[ r \left( \Erf\left(\frac{r}{\sqrt{2}}\right)^2-1\right)\right] \\
   &=& \left( \frac{2}{\pi} \right)^{3/2} \int_{r=0}^\infty K_0\left( \frac{r^2}{2}\right) \Erf\left(\frac{r}{\sqrt{2}}\right) r \dd r. \nonumber
\end{eqnarray}
In the second line of~\eqref{I-app-3} the second term vanishes and the derivative of the $r$ part in the third term exactly cancels the first term, 
so that only the term on the last line remains. Finally, we use the definition of $\Erf$ and an integral representation of the $K_0$ function,  
\begin{eqnarray}
\label{I-app-4}
I&=& \frac{4 \sqrt{2}}{\pi^2} \int_{r=0}^\infty \int_{v=0}^\frac{r}{\sqrt{2}} \ee^{-v^2} 
\dd v \int_{t=1}^\infty \frac{\ee^{-\frac{r^2 t}{2}}}{\sqrt{t^2-1}} \dd t r \dd r \nonumber \\
&=&  \frac{4 \sqrt{2}}{\pi^2} \int_{t=1}^\infty \frac{\dd t}{\sqrt{t^2-1}} \int_{v=0}^\infty \dd v \ee^{-v^2} \int_{u=v^2}^\infty \ee^{-u t} \dd u \\
&=& \frac{2 \sqrt{2}}{\pi^{3/2}} \int_{t=1}^\infty \frac{\dd t}{t(t+1)\sqrt{t-1}} = \frac{2}{\sqrt{\pi}} (\sqrt{2}-1).\nonumber
\end{eqnarray}
In equation~\eqref{I-app-4}, after expressing the $\Erf$ and $K_0$ functions, we made the change of variables $u = \frac{r^2}{2}$, 
then performed the integrals over $u$, $v$ and $t$ in that order. We get $I = \frac{2}{\sqrt{\pi}} (\sqrt{2}-1)$ as announced in the main text.

\end{document}